\documentclass[journal]{new-aiaa}
\usepackage[utf8]{inputenc}

\usepackage{textcomp}
\usepackage{graphicx}%

\usepackage{bm}
\usepackage{amsmath}
\usepackage{xcolor}%

\usepackage{colortbl}
\definecolor{myblue}{rgb}{0, 0.23, 0.64}
\definecolor{WVUblue}{rgb}{0, 0.16, 0.33}
\definecolor{gray_row}{HTML}{F0F0F0}

\usepackage{hyperref}
\hypersetup{
    colorlinks=true,
    linkcolor=myblue,
    filecolor=magenta,
    citecolor=myblue,
    urlcolor=myblue,
}

\usepackage[version=4]{mhchem}
\usepackage{siunitx}
\usepackage{longtable,tabularx}
\usepackage{gensymb}
\usepackage{mathtools}
\usepackage{float}

\usepackage{subcaption}
\usepackage{enumitem}

\usepackage{multirow}
\usepackage{booktabs}
\usepackage{pdflscape}
\usepackage{mdframed}
\usepackage[flushleft]{threeparttable}
\usepackage{multirow}
\usepackage{xspace}
\usepackage{cancel}

\let\svthefootnote\thefootnote
\newcommand\freefootnote[1]{%
  \let\thefootnote\relax%
  \footnotetext{#1}%
  \let\thefootnote\svthefootnote%
}

\mdfsetup{
middlelinecolor=red,
middlelinewidth=2pt,
backgroundcolor=red!5}

\title{Venusian Radio Occultation Constellation Design and Reconfiguration Problems: A Benders Decomposition Approach}

\author{David O. Williams Rogers\footnote{Ph.D. Student, Department of Mechanical, Materials and Aerospace Engineering, Student Member AIAA.}, Trupti P. Gosavi \footnote{Ph.D. Student, Department of Mechanical, Materials and Aerospace Engineering, Student Member AIAA.}, and Hang Woon Lee\footnote{Assistant Professor, Department of Mechanical, Materials and Aerospace Engineering; hangwoon.lee@mail.wvu.edu. Member AIAA (Corresponding Author).}}
\affil{West Virginia University, Morgantown, West Virginia, 26506}

\begin{document}

\freefootnote{This paper is a substantially revised and expanded version of the Paper AAS 24-395, presented at the 2024 AAS/AIAA Astrodynamics Specialist Conference, Bloomfield, CO, August 11-15, 2024. It offers new formulations, solution methods, numerical experiments, and a better description of the materials.}

\maketitle

\begin{abstract}
Driven by the scientific value of Venusian atmospheric studies, numerous missions, such as Magellan and Venus Express, have relied on single-spacecraft radio occultation (RO) experiments. Despite their proven success, data throughput is inherently limited by the use of a single transmitter and its relative geometry with respect to Earth-based receivers. Therefore, the literature proposes leveraging a constellation of spacecraft to increase RO measurement opportunities and improve spatial and temporal coverage; however, current efforts impose strict simplifying assumptions that limit their applicability and responsiveness to complex mission narratives. Beyond initial constellation design, evolving mission objectives may require event-based constellation reconfiguration during operations. To address both design and operational needs, we propose two problems formulated as integer linear programs: the RO Constellation Configuration Design Problem and the RO Constellation Reconfiguration Problem. Further, we address the programs' solution-space explosion with a Benders decomposition approach. Numerical experiments show the applicability of our problems for constellation configuration design, tackle the full mission pipeline, transitioning from global atmospheric profiling to on-demand regional profiling above areas of volcanic interest, and show that the Benders decomposition approach efficiently obtains high-quality solutions to large-scale instances.
\end{abstract}

\section*{Nomenclature}
{\renewcommand\arraystretch{1.0}
\noindent\begin{longtable*}{@{}l @{\quad=\quad} l@{}}
\multicolumn{2}{@{}l}{Sets}\\
$\mathcal{J}$ & orbital slots\\
$\mathcal{K}$ & time steps\\
$\mathcal{O}$ & atmospheric regions\\
$\mathcal{P}$ & spacecraft pairs\\
$\mathcal{S}$ & spacecraft\\
\multicolumn{2}{@{}l}{Subscripts and indexing}\\
$i,j$ & orbital slots\\
$k$ & time step \\
$p$ & spacecraft pair\\
$o$ & atmospheric region\\
$s$ & spacecraft\\
\multicolumn{2}{@{}l}{Parameters}\\
$\bm{c}$ & orbital transfer $\Delta v$ cost\\
$\bm{C}$ & Benders optimality cuts\\
$S$ & spacecraft cardinality\\
$\bm{V}$ & radio occultation measurement feasibility\\
$\bm{\varepsilon}$ & $\Delta v$ transfer budget\\
$\bm{\pi}$ & radio occultation measurement reward\\
\multicolumn{2}{@{}l}{Decision variables}\\
$\bm{w}$ & orbital transfer\\
$\bm{x}$ & spacecraft location\\
$\bm{y}$ & spacecraft pair\\
$\bm{z}$ & radio occultation measurement\\
$\bm{\alpha}$ & subproblem dual variables\\
$\bm{\beta}$ & subproblem dual variables\\
$\theta$ & auxiliary continuous variable\\
\end{longtable*}}

\section{Introduction}\label{sec:introduction}
Characterizing habitable environments, from their origin to their evolution, is a key scientific question \cite{decadal_nasem_2023} that can be addressed by conducting comparative studies across planets \cite{kane_agu_2019}. Among the eight planets in the solar system, Venus is the most comparable to Earth, differing by \SI{18.50}{\%} in mass and \SI{2}{\%} in mean density (reduced to the same pressure) \cite{kaula_science_1990}. In addition, it is assumed that the two planets shared similarities in their origins, including volatile and refractory compound compositions; however, unlike Earth, Venus transitioned from a habitable to an uninhabitable environment \cite{arney_pa_2020}. 

Among the many factors that influence habitability, the planetary atmosphere is key \cite{kane_nature_2024}; thus, characterizing its composition, structure, and dynamics is essential for advancing our understanding. To achieve this characterization, a range of remote sensing methods is utilized, including infrared \cite{marcq_pss_2006} and Doppler \cite{sanchez_icarus_2017} spectroscopy, and radio occultation (RO) \cite{jenkins_icarus_1994}. Within this cohort, RO is of particular interest as its measurements enable the derivation of temperature and pressure profiles in the neutral atmosphere, as well as vertical profiles of sulfuric acid \cite{jenkins_icarus_1994}.

Atmospheric RO measurements involve a receiver (\textit{e.g.,} an Earth-based ground station) acquiring radio waves from a transmitter (\textit{e.g.,} a spacecraft), which is occulted by the planetary body of interest and its atmosphere. The transmitted waves experience bending, attenuation, and scintillation as a consequence of their interaction with the planet's atmosphere \cite{imamura_eps_2011}. These perturbations are induced by atmospheric refraction and are used to infer the physical properties of the medium, a process commonly known as the inverse problem \cite{bourgoin_aa_2029}. For example, using geometric optics, the refractive index is obtained as a function of the ray's closest approach to the planet (commonly referred to as the RO tangent point) and its bending angle \cite{fjeldbo_aj_1971}, where the ray bends towards higher refractivity regions \cite{caruso_rs_2023}. This index is useful because it enables the determination of temperature and pressure profiles \cite{eshleman_pss_1973} and the relative volume content of sulfuric acid \cite{gubenko_cosmic_2001}. Successful missions that collected the data required to derive such profiles include, but are not limited to, the Magellan spacecraft \cite{jenkins_icarus_1994} and the Akatsuki orbiter \cite{imamura_eps_2017}.

The scientific data throughput from Venusian RO missions can be maximized by adopting a constellation instead of a single-transmitter spacecraft. Essentially, rather than relying on Earth-based ground receivers, the RO measurements can be conducted using pairs of Venus-orbiting spacecraft, adopting a concept of operations similar to Earth-based RO, where Global Navigation Satellite Systems (GNSS) are used as a transmitter and a spacecraft constellation acts as a receiver (\textit{e.g.,} Global Positioning System and the FORMOSAT-3/COSMIC \cite{yue_sw_2014}). In addition, the adoption of constellations increases mission robustness to failures and offers the opportunity to lower life-cycle costs. However, enhancements in scientific data throughput depend on complex factors that extend beyond simply increasing the number of spacecraft. In particular, the \textit{constellation configuration} governs the inter-spacecraft geometry, thereby serving as a primary driver of the RO measurements, both in terms of quality, total number, and atmospheric location of the tangent point.

The literature addresses RO constellation configuration design by adopting several simplifying assumptions. For example, Refs.~\cite{juang_aa_2013,Asgarimehr_jg_2015,xu_rs_2019} focus on receiver spacecraft placement assuming a preexisting GNSS constellation as the transmitting spacecraft, and Ref.~\cite{chandratre_scitech_2024} designs an RO constellation configuration for a three-6U-CubeSat constellation with fixed payload characteristics. Stepping back a little and considering the neighboring literature, many satellite constellation design methods have been proposed for remote sensing, space situational awareness, and other applications (see Refs.~\cite{lee_jsr_2020,lee_jsr_2023,williams_jsr_2026}). However, these methods are not directly applicable to RO constellation design because they are built around a fundamentally different notion of coverage. In conventional applications, coverage is determined by whether an individual satellite can observe a target. In RO, by contrast, coverage is determined by a transmitter-receiver satellite pair whose geometry produces a tangent point in the atmosphere.

Despite the promising approaches proposed in the literature to address the RO constellation configuration design problem, their applicability to Venusian RO missions presents certain limitations. First, they enforce strict modeling assumptions, which significantly constrain the design space and limit their applicability to particular mission scenarios and planetary bodies (\textit{e.g.,} the Earth). Second, existing efforts design RO constellation configurations assuming full \textit{a priori} knowledge of the mission scenario. Consequently, because they do not account for unplanned changes in mission objectives during their design, their performance may degrade significantly when such objectives change suddenly. For instance, after the constellation's deployment, operators may prioritize local RO atmospheric measurements near recently-detected volcanic activity to study changes in atmospheric composition \cite{esposito_science_1984}, rather than targeting global atmospheric studies.

\textit{Constellation reconfiguration}, the process of achieving a new configuration through spacecraft performing orbital transfers, offers a practical resolution for constellation responsiveness to unplanned planetary phenomena. In the context of Earth observation mission planning, the literature demonstrates the benefits of adopting reconfiguration to address localized, time-sensitive events. Reference~\cite{lee_jsr_2023} proposes a formulation to reconfigure the constellation, obeying $\Delta v$ transfer budgets, to maximize the time-dependent observations supporting post-disaster operations over regions of interest. Similarly, Ref.~\cite{lee_jsr_2025} determines an optimal sequence of constellation reconfigurations that maximizes the observations over highly dynamic targets (\textit{e.g.,} hurricanes), and demonstrates the performance improvements over a non-reconfigurable constellation. 

Although the literature demonstrates the value of adopting constellation reconfiguration to maximize mission performance when mission objectives change or are highly dynamic, existing approaches cannot be directly extrapolated to our problem domain. In particular, current efforts target remote sensing applications that adopt coverage as the key figure of merit. As a result, they fail to capture the required inter-satellite geometry and RO tangent point location.

\begin{figure}[htpb]
    \centering
    \includegraphics[width=\linewidth]{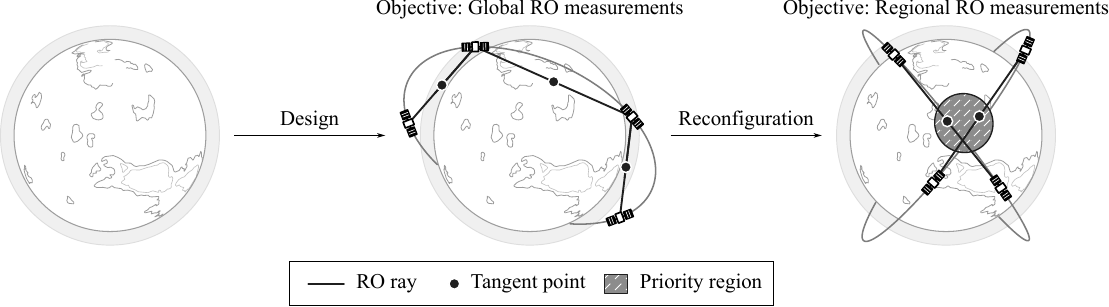}
    \caption{RO-CDP addresses constellation configuration design, and RO-CRP determines the optimal constellation reconfiguration.}
    \label{fig:overarching}
\end{figure}

This article addresses the literature's limitations by proposing two problems formulated as integer linear programs (ILPs) to underpin the design and reconfiguration optimization of Venusian RO constellations. The two problems are obtained by tailoring the Regional Constellation Reconfiguration Problem (RCRP) \cite{lee_jsr_2023} (\textit{i.e.,} concurrent constellation configuration design and constellation reconfiguration problems for Earth observation applications, formulated as ILPs) for RO applications. The first problem, the \textit{RO Constellation Configuration Design Problem} (RO-CDP), determines the locations of a user-defined number of spacecraft to maximize the time-dependent RO measurement rewards over the atmospheric regions of interest. The second problem, the \textit{RO Constellation Reconfiguration Problem} (RO-CRP), determines, on a per-event or mission-horizon basis, the constellation reconfiguration that maximizes the time-dependent RO measurement rewards over the atmospheric regions of interest, subject to predefined spacecraft $\Delta v$ transfer budgets. Figure~\ref{fig:overarching} illustrates the initial design of a constellation configuration to achieve global RO atmospheric measurements using RO-CDP, and subsequently, given a change in the mission objective, the constellation reconfigures to conduct regional RO measurements over a defined priority atmospheric region using RO-CRP. Recognizing the combinatorial nature of the two problems' formulations, we propose a Benders decomposition approach to obtain high-quality solutions for large-scale instances, which would otherwise be challenging due to the explosion of their solution space. This paper extends a previous version of this research \cite{trupti_aas_2024} by providing new formulations, solution methods, and numerical experiments.

The remainder of this article is organized as follows. Section~\ref{sec:methodology} introduces the proposed methodology, Sec.~\ref{sec:numerical_experiments} presents the numerical experiments, and Sec.~\ref{sec:conclusions} concludes the article.

\section{Methodology}\label{sec:methodology}
RO-CDP and RO-CRP address constellation configuration design and reconfiguration optimization, respectively, for Venus atmospheric studies using RO. These two problems are inspired by RCRP \cite{lee_jsr_2023}, which is formulated as an ILP that concurrently tackles constellation configuration design and reconfiguration for complex regional coverage. Specifically, RCRP comprises two problems: the constellation configuration design problem, which determines the optimal orbital distribution of a fixed number of spacecraft, and the constellation reconfiguration problem, which determines the optimal constellation reconfiguration.

Both RO-CDP and RO-CRP are formulated as ILPs and share common discrete sets. First, the set of time steps $\mathcal{K}$, with index $k$ and cardinality $K$, which discretizes the mission duration. Second, the set of orbital slots $\mathcal{J}$, with index $j$ and cardinality $J$, where each element is associated with a set of orbital elements defined at the mission epoch. Third, the set of spacecraft pairs $\mathcal{P}$, with index $p$ and cardinality $P$, which is constructed from all possible combinations between orbital slots. Namely, each $p$ has an associated set $\mathcal{J}_{p}$ that contains its corresponding pair of orbital slots $(i,j) \in \mathcal{J}$. Fourth, the set of atmospheric regions $\mathcal{O}$, with index $o$ and cardinality $O$. Each atmospheric region $o \in \mathcal{O}$ is a polyhedron, with vertices specified by geodetic latitude, longitude, and altitude.

Similarly, both problems share a common set of binary decision variables. First, the set of spacecraft pair decision variables $\bm{y}\in\{0,1\}^{P}$, where each of its elements is defined as:
\begin{equation}
    y_{p} = \begin{cases}
        1, &\text{if spacecraft's pair $p$ is selected}\\
        0, &\text{otherwise}
    \end{cases}
\end{equation}
Second, the set of RO measurement decision variables $\bm{z}\in\{0,1\}^{K\times O}$, where each element is defined as:
\begin{equation}
    z_{ko} = \begin{cases}
        1, &\text{if at time step $k$, an RO measurement is taken at atmospheric region $o$}\\
        0, &\text{otherwise}
    \end{cases}
\end{equation}
In addition, we define for time step $k$ and atmospheric region $o$ the RO measurement reward $\pi_{ko} \in \mathbb{R}_{\ge 0}$. The definition of this reward parameter enables mission designers to weigh particular atmospheric regions during distinct mission phases. For example, it provides flexibility to define global or regional measurement objectives.

To encode whether an atmospheric region is measured through RO (\textit{i.e.,} the transmitted wave's tangent point lies within that atmospheric region), we introduce parameter $\bm{V}\in\{0,1\}^{K\times P\times O}$, where each element is defined as:
\begin{equation}
    V_{kpo}=\begin{cases}
        1, & \text{if at time step $k$, an RO measurement between spacecraft pair $p$ over atmospheric region $o$ occurs}\\
        0, & \text{otherwise}
    \end{cases}
\end{equation}
The definition of this parameter is sufficiently flexible to allow mission designers to compute the locations of the RO tangent points as desired (\textit{e.g.,} analytical ray-tracing \cite{caruso_rs_2023}). 

Lastly, Table~\ref{tab:common_params} summarizes the common sets, parameters, and decision variables used by both problems.
\begin{table}[htpb]
    \caption{Common sets, parameters, and decision variables.}
    \label{tab:common_params}
    \centering
    \begin{tabular}{lll}
        \hline
        \hline
        Type & Symbol & Description\\
        \hline
             Sets   & $\mathcal{J}$ & Set of orbital slots (index $j$; cardinality $J$) \\
                    & $\mathcal{K}$ & Set of time steps (index $k$; cardinality $K$) \\
                    & $\mathcal{O}$ & Set of atmospheric regions (index $o$; cardinality $O$) \\
                    & $\mathcal{P}$ & Set of spacecraft pairs (index $p$; cardinality $P$) \\
             Parameters & $\pi_{ko}$ & RO measurement reward\\ 
                    & $S$ & Number of spacecraft in the constellation\\
                    & $V_{kpo}$ & $\begin{cases}
                                        1, & \text{if at time step $k$, an RO measurement between spacecraft pair $p$ over}\\
                                        &\text{atmospheric region $o$ occurs}\\
                                        0, & \text{otherwise}
                                    \end{cases}$ \\
            Decision variables & $y_{p}$&$ \begin{cases}
                                            1, &\text{if spacecraft's pair $p$ is selected}\\
                                            0, &\text{otherwise}
                                        \end{cases}$\\
                                & $z_{ko}$ & $\begin{cases}
                                            1, &\text{if at time step $k$, an RO measurement is taken at atmospheric region $o$}\\
                                            0, &\text{otherwise}
                                        \end{cases}$\\
         \hline
         \hline
    \end{tabular}
\end{table}

\subsection{Radio Occultation Constellation Configuration Design Problem}\label{sec:methodology_cdp}
RO-CDP determines the location of the $S$ spacecraft that maximizes the RO measurement rewards. In addition to the two sets of decision variables presented before, RO-CDP uses the set of location decision variables $\bm{x} \in \{0,1\}^{J}$, where each element is defined as:
\begin{equation}
    x_j = \begin{cases}
        1, &\text{if orbital slot $j$ is occupied by a spacecraft}\\
        0, &\text{otherwise}
    \end{cases}
\end{equation}

Then, RO-CDP reads as follows:
\begin{subequations}
    \begin{alignat}{2}
        \text{(RO-CDP)} \quad \max \quad &\sum_{k \in \mathcal{K}}\sum_{o \in \mathcal{O}}\pi_{ko}z_{ko} \label{rocdp:obj}\\
        \text{s.t.} \quad &\sum_{p \in \mathcal{P}}V_{kpo}y_{p} \ge z_{ko}, &\quad \forall k \in \mathcal{K}, \forall o \in \mathcal{O} \label{rocdp:visibility} \\
        &y_{p} \le \frac{1}{2}(x_i + x_j), &\quad \forall p \in \mathcal{P}, \forall (i,j) \in \mathcal{J}_{p} \label{rocdp:activation}\\
        & \sum_{j \in \mathcal{J}}x_j=S \label{rocdp:cardinality}\\
        &x_{j} \in \{0,1\}, &\quad \forall j \in \mathcal{J}\label{rocdp:dv_x}\\
        &y_{p} \in \{0,1\}, &\quad \forall p \in \mathcal{P}\label{rocdp:dv_y}\\
        &z_{ko} \in \{0,1\}, &\quad \forall k \in \mathcal{K}, \forall o \in \mathcal{O}\label{rocdp:dv_z}
    \end{alignat}
\end{subequations}
Objective function~\eqref{rocdp:obj} maximizes the RO measurement rewards. For each time step, constraints~\eqref{rocdp:visibility} encode feasible RO measurement rewards in the defined atmospheric regions. Constraints~\eqref{rocdp:activation} activate a spacecraft pair only if the corresponding orbital slots are occupied. Furthermore, constraints~\eqref{rocdp:cardinality} enforce the number of spacecraft in the constellation, and constraints~\eqref{rocdp:dv_x},~\eqref{rocdp:dv_y}, and~\eqref{rocdp:dv_z} define the decision variables' binary domain.

\subsection{Radio Occultation Constellation Reconfiguration Problem}\label{sec:methodology_crp}
RO-CRP determines the single-stage constellation reconfiguration that maximizes the RO measurement rewards. To formulate this problem, we define the set of spacecraft in the constellation $\mathcal{S}$, with index $s$ and cardinality $S$. Then, we introduce the cost parameter $c_{sj} \in \mathbb{R}_{\ge0}$ to encode the required $\Delta v$ for spacecraft $s$ to transfer from its initial orbital slot to its final orbital slot $j$. Note that a cost equal to zero indicates that a platform maintains its initial orbital slot.

To model the spacecraft's orbital transfers, we propose the set of orbital transfer decision variables $\bm{w} \in \{0,1\}^{S\times J}$, where each element is defined as: 
\begin{equation}
    w_{sj} = \begin{cases}
        1, & \text{if spacecraft $s$ transfers to orbital slot $j$}\\
        0, & \text{otherwise}
    \end{cases}
\end{equation}

Then, RO-CRP is formulated as follows: 
\begin{subequations}
    \begin{alignat}{2}
        \text{(RO-CRP)} \quad \max \quad &\sum_{k \in \mathcal{K}}\sum_{o \in \mathcal{O}}\pi_{ko}z_{ko} \tag{\ref{rocdp:obj}}\\
        \text{s.t.} \quad &\sum_{p \in \mathcal{P}}V_{kpo}y_{p} \ge z_{ko}, &\quad \forall k \in \mathcal{K}, \forall o \in \mathcal{O} \tag{\ref{rocdp:visibility}} \\
        &y_{p} \le \frac{1}{2}\sum_{s \in \mathcal{S}}(w_{si} + w_{sj}), &\quad \forall p \in \mathcal{P}, \forall (i,j) \in \mathcal{J}_{p} \label{crp:activation}\\
        &\sum_{s \in \mathcal{S}}w_{sj}\le1, & \quad \forall j \in \mathcal{J}  \label{crp:assingment_slot}\\
        &\sum_{j \in \mathcal{J}}w_{sj}=1, & \quad \forall s \in \mathcal{S} \label{crp:assingment_sat}\\
        &\sum_{j \in \mathcal{J}}c_{sj}w_{sj} \le \varepsilon_s, & \quad \forall s \in \mathcal{S} \label{crp_e:dv}\\
        &w_{sj} \in \{0,1\}, &\quad \forall s \in \mathcal{S},\forall j \in \mathcal{J}\label{crp:dv_w}\\
        &y_{p} \in \{0,1\}, &\quad \forall p \in \mathcal{P}\tag{\ref{rocdp:dv_y}}\\
        &z_{ko}\in \{0,1\},&\quad \forall k \in \mathcal{K}, \forall o \in \mathcal{O} \tag{\ref{rocdp:dv_z}}
    \end{alignat}
\end{subequations}
Constraints~\eqref{crp:activation} require that a spacecraft pair is selected if the corresponding destination orbital slots are occupied. Constraints~\eqref{crp:assingment_slot} enforce that an orbital slot can be occupied by at most one spacecraft, and constraints~\eqref{crp:assingment_sat} require each spacecraft to occupy exactly one orbital slot. Furthermore, constraints~\eqref{crp_e:dv} impose a per-spacecraft $\Delta v$ transfer budget, denoted by parameter $\varepsilon_s \in \mathbb{R}_{\ge0}$. Lastly, constraints~\eqref{crp:dv_w} define the binary domain of the orbital transfer decision variables.

\subsection{Branch-and-Benders-Cut Algorithm}
\textit{Benders decomposition} is a solution method for mixed-integer linear programming (MILP) problems that can be partitioned into a master problem containing the complicating variables and a linear subproblem obtained by fixing those variables \cite{benders}. The dual of the linear subproblem is used to generate linear constraints, called \textit{Benders cuts}, for the master problem. The benefits of this decomposition method for large-scale, complex problems have been demonstrated in applications including power system optimization \cite{shahidehopour_ieee_2005}, logistics \cite{adulyasak_or_2015}, and network design problems \cite{botton_informs_2013}. 

There are two types of cuts that can be generated from the dual subproblem. First, feasibility cuts exclude master decisions for which the primal subproblem is infeasible. If the dual subproblem is unbounded below, weak duality implies that the primal subproblem is infeasible \cite{bertsimas_book}; an improving extreme ray of the dual polyhedron can then be used to generate a feasibility cut. Second, optimality cuts help tighten the bounds of the master problem. When the primal subproblem is feasible and bounded, an optimal dual solution can be used to generate an optimality cut. Such a solution can be selected at an extreme point of the dual polyhedron.

Benders decomposition is typically implemented following two methods: the cutting planes procedure and the branch-and-Benders-cut (B\&BC) algorithm. While the former method repeatedly solves the master problem after adding cuts, the latter generates Benders cuts within a single branch-and-bound search of the master problem. A substantial share of the state-of-the-art literature (\textit{e.g.,} Refs.~\cite{huang_ieee_2019,li_cor_2020,castelluci_esa_2025}) relies on the latter method, given that it solves the master problem only once, and that MILP techniques (\textit{e.g.,} primal heuristics, preprocessing) can be combined with the algorithm more efficiently \cite{cordeau_ejor_2019}.  

To apply the B\&BC algorithm to RO-CDP and RO-CRP, we must select an appropriate decomposition into a master problem and a subproblem, ensuring that the latter is a linear program. Although both problems are formulated as ILPs with binary decision variables, their structure allows us to obtain such a subproblem by relaxing the binary domain constraints on the RO measurement decision variables (constraints~\eqref{rocdp:dv_z}) to permit values in the closed interval from zero to one. Specifically, if the left-hand side of constraints~\eqref{rocdp:visibility} is zero, then the relaxed RO measurement decision variables are forced to be zero. Conversely, if the left-hand side of these constraints is greater than or equal to one, the relaxed RO measurement decision variables can take the value of one to maximize the value of objective function~\eqref{rocdp:obj}. Accordingly, we replace the binary domain constraints~\eqref{rocdp:dv_z} with the bounds specified by constraints~\eqref{sp:z_ub} and~\eqref{sp:z_lb}:
\begin{subequations}
    \begin{alignat}{2}
        &z_{ko} \le 1, &\quad \forall k \in \mathcal{K}, \forall o \in \mathcal{O}\label{sp:z_ub}\\
        &z_{ko} \ge 0, &\quad \forall k \in \mathcal{K}, \forall o \in \mathcal{O}\label{sp:z_lb}
    \end{alignat}
\end{subequations}

\subsubsection{Benders Decomposition for RO-CDP}
Projecting out the relaxed RO measurement decision variables, the master problem for RO-CDP selects the constellation configuration and spacecraft pairs, and reads as:
\begin{subequations}
\label{MP:loc_formulation}
    \begin{alignat}{2}
        \text{(Master for RO-CDP)} \quad \max \quad &\theta \label{MP:obj}\\
        \text{s.t.} \quad &y_{p} \le \frac{1}{2}(x_i + x_j), &\quad \forall p \in \mathcal{P}, \forall (i,j) \in \mathcal{J}_{p} \tag{\ref{rocdp:activation}}\\
        & \sum_{j \in \mathcal{J}}x_j=S \tag{\ref{rocdp:cardinality}}\\
        &\theta \le C_{e}(\bm{y}), &\quad \forall e \in \mathcal{E} \label{mp:cuts}\\
        & \theta \le \sum_{k\in\mathcal{K}}\sum_{o\in\mathcal{O}} \pi_{ko}\label{mp:init_bound}\\
        & \theta \ge 0 \label{mp:theta_lb}\\
        &x_{j} \in \{0,1\}, &\quad \forall j \in \mathcal{J}\tag{\ref{rocdp:dv_x}}\\
        &y_{p} \in \{0,1\}, &\quad \forall p \in \mathcal{P}\tag{\ref{rocdp:dv_y}}
    \end{alignat}
\end{subequations}
Here, $\theta$ is a nonnegative auxiliary variable representing the measurement reward in the master problem. Until sufficient optimality cuts have been added, the master problem may overestimate the reward achievable by its selected configuration. Let $\mathcal{E}$, indexed by $e$, denote the set of generated optimality cuts. Constraints~\eqref{mp:cuts} restrict this estimate, while constraints~\eqref{mp:init_bound} and \eqref{mp:theta_lb} provide initial bounds.

Once the master problem is defined, the RO-CDP primal subproblem for any fixed $(\bar{\bm{x}},\bar{\bm{y}})$ evaluates the resulting RO measurement rewards, and reads as:
\begin{subequations}\label{benders:sp}
    \begin{alignat}{2}
        \text{(Primal Subproblem)} \quad \max \quad &\sum_{k \in \mathcal{K}}\sum_{o \in \mathcal{O}}\pi_{ko}z_{ko}\tag{\ref{rocdp:obj}}\\
        \text{s.t.} \quad & z_{ko} \le \sum_{p \in \mathcal{P}}V_{kpo}\bar{y}_{p}, &\quad \forall k \in \mathcal{K}, \forall o \in \mathcal{O} \label{sp:visibility}\\
        &z_{ko} \le 1, &\quad \forall k \in \mathcal{K}, \forall o \in \mathcal{O}\tag{\ref{sp:z_ub}}\\
        &z_{ko} \ge 0, &\quad \forall k \in \mathcal{K}, \forall o \in \mathcal{O}\tag{\ref{sp:z_lb}}
    \end{alignat}
\end{subequations}
where constraints \eqref{sp:visibility} activate the relaxed RO measurement decision variables only if at least one of the fixed spacecraft pair has a tangent point lying on that atmospheric region. 

Then, the dual subproblem is given as:
\begin{subequations}\label{benders:sp_dual}
    \begin{alignat}{2}
        \text{(Dual Subproblem)} \quad \min \quad &\sum_{k \in \mathcal{K}}\sum_{o \in \mathcal{O}}\alpha_{ko}\sum_{p \in \mathcal{P}}V_{kpo}\bar{y}_{p} + \sum_{k \in \mathcal{K}}\sum_{o \in \mathcal{O}}\beta_{ko} \label{dual:obj}\\
        \text{s.t.} \quad &\alpha_{ko}+\beta_{ko} \ge \pi_{ko}, &\quad \forall k \in \mathcal{K}, \forall o \in \mathcal{O} \label{dual:coupling} \\
        &\alpha_{ko},\beta_{ko} \ge 0, &\quad \forall k \in \mathcal{K}, \forall o \in \mathcal{O}\label{sp:z}
    \end{alignat}
\end{subequations}
where $\bm{\alpha} \in \mathbb{R}_{\ge0}^{K\times O}$, with elements $\alpha_{ko}$, are the dual variables associated with constraints~\eqref{sp:visibility}, and $\bm{\beta}\in \mathbb{R}_{\ge0}^{K\times O}$, with elements $\beta_{ko}$, are the dual variables associated with constraints~\eqref{sp:z_ub}. Note that the dual subproblem cannot be unbounded, since the dual variables are bounded from below. Thus, no feasibility cuts are derived from the dual. Then, given an extreme point $e$ with optimal decision variables $(\bar{\bm{\alpha}},\bar{\bm{\beta}})$, Benders optimality cuts are formulated as:
\begin{equation}
    C_e(\bm{y})=\sum_{k \in \mathcal{K}}\sum_{o \in \mathcal{O}}\bar{\alpha}_{ko}\sum_{p \in \mathcal{P}}V_{kpo}y_{p} + \sum_{k \in \mathcal{K}}\sum_{o \in \mathcal{O}}\bar{\beta}_{ko}
\end{equation}

\subsubsection{Benders Decomposition for RO-CRP}
The master problem for RO-CRP selects the constellation reconfiguration and corresponding spacecraft pairs, and is obtained by projecting out the relaxed RO measurement decision variables. Then, the master problem reads as:
\begin{subequations}
\label{MP:rec_formulation}
    \begin{alignat}{2}
        \text{(Master for RO-CRP)} \quad \max \quad &\theta \tag{\ref{MP:obj}}\\
        \text{s.t.} \quad 
        &y_{p} \le \frac{1}{2}\sum_{s \in \mathcal{S}}(w_{si} + w_{sj}), &\quad \forall p \in \mathcal{P}, \forall (i,j) \in \mathcal{J}_{p} \tag{\ref{crp:activation}}\\
        &\sum_{s \in \mathcal{S}}w_{sj}\le1, & \quad \forall j \in \mathcal{J}  \tag{\ref{crp:assingment_slot}}\\
        &\sum_{j \in \mathcal{J}}w_{sj}=1, & \quad \forall s \in \mathcal{S} \tag{\ref{crp:assingment_sat}}\\
        &\sum_{j \in \mathcal{J}}c_{sj}w_{sj} \le \varepsilon_s, &\quad \forall s \in \mathcal{S} \tag{\ref{crp_e:dv}}\\
        & \theta \le C_{e}(\bm{y}), &\quad \forall e \in \mathcal{E} \tag{\ref{mp:cuts}}\\
        & \theta \le \sum_{k\in\mathcal{K}}\sum_{o\in\mathcal{O}} \pi_{ko}\tag{\ref{mp:init_bound}}\\
         & \theta \ge 0 \tag{\ref{mp:theta_lb}}\\
        &w_{sj} \in \{0,1\}, &\quad \forall s \in \mathcal{S},\forall j \in \mathcal{J}\tag{\ref{crp:dv_w}}\\
        &y_{p} \in \{0,1\}, &\quad \forall p \in \mathcal{P}\tag{\ref{rocdp:dv_y}}
    \end{alignat}
\end{subequations}

Although the RO-CRP and RO-CDP master problems differ, their subproblems have the same structure. Consequently, the same dual-subproblem formulation and optimality-cut construction apply to both. Numerical cut coefficients depend on the visibility and reward data of the particular instance and on the dual solution used to generate each cut.

\section{Numerical Experiments}\label{sec:numerical_experiments}

This section presents a set of numerical experiments that demonstrate the value of the contributions of this article. First, Sec.~\ref{sec:numexp_benders_cd} and Sec.~\ref{sec:numexp_benders_cr} benchmark the proposed B\&BC algorithm for RO-CDP and RO-CRP. Second, Sec.~\ref{sec:numexp_cd_cr} shows the applicability of both ILPs to tackle the full mission design and operation pipeline to study Venus' atmosphere, considering dynamic mission objectives. Lastly, Sec.~\ref{sec:mission_design} performs a sensitivity analysis between the constellation configuration design and the regional, time-dependent RO measurement rewards definition.

All numerical experiments share a set of underlying assumptions and are solved to optimality (otherwise stated) using the Gurobi Optimizer 13.0.0 with default settings on an Apple M4 Max. The first underlying assumption is that Venus is represented as a perfect sphere, based on the magnitudes (\textit{i.e.,} order of \num{e-6}) of the harmonic coefficients $J_{2}$, $J_{3}$, and $J_{4}$, which enables us to propagate the orbits using the two-body equations of motion. The second underlying assumption states that the $z$-axis of the Venus-Centered Inertial (VCI) coincides with the Venus-Centered Venus-Fixed frame. At the mission epoch, the $x$-axis of both frames is aligned with the Ariadne crater, which serves as the reference point for the prime meridian. The third underlying assumption posits that the location of an RO measurement is approximated by the tangent point of a sphere centered at Venus' center of mass with the vector connecting the spacecraft pair. This assumption stems from the small bending angles (\textit{i.e.,} less than \SI{2}{deg}) at altitudes above \SI{50}{km} \cite{imamura_eps_2017}.

\subsection{Branch-and-Benders-Cut RO Constellation Configuration Design Benchmark}\label{sec:numexp_benders_cd}

This benchmark study demonstrates the benefits of using B\&BC compared to using Gurobi Optimizer to solve the baseline model (\textit{i.e.,} Sec.~\ref{sec:methodology_cdp}). 

\subsubsection{Setup}

For this benchmark study, we modify two of the Gurobi Optimizer's default settings. First, we set the runtime limit to at most four hours. Second, we enable lazy constraints to implement B\&BC using the Gurobi Optimizer.

\begin{figure}[htpb]
    \centering
    \begin{subfigure}[c]{0.44\linewidth}
        \centering
        \includegraphics[width=\linewidth]{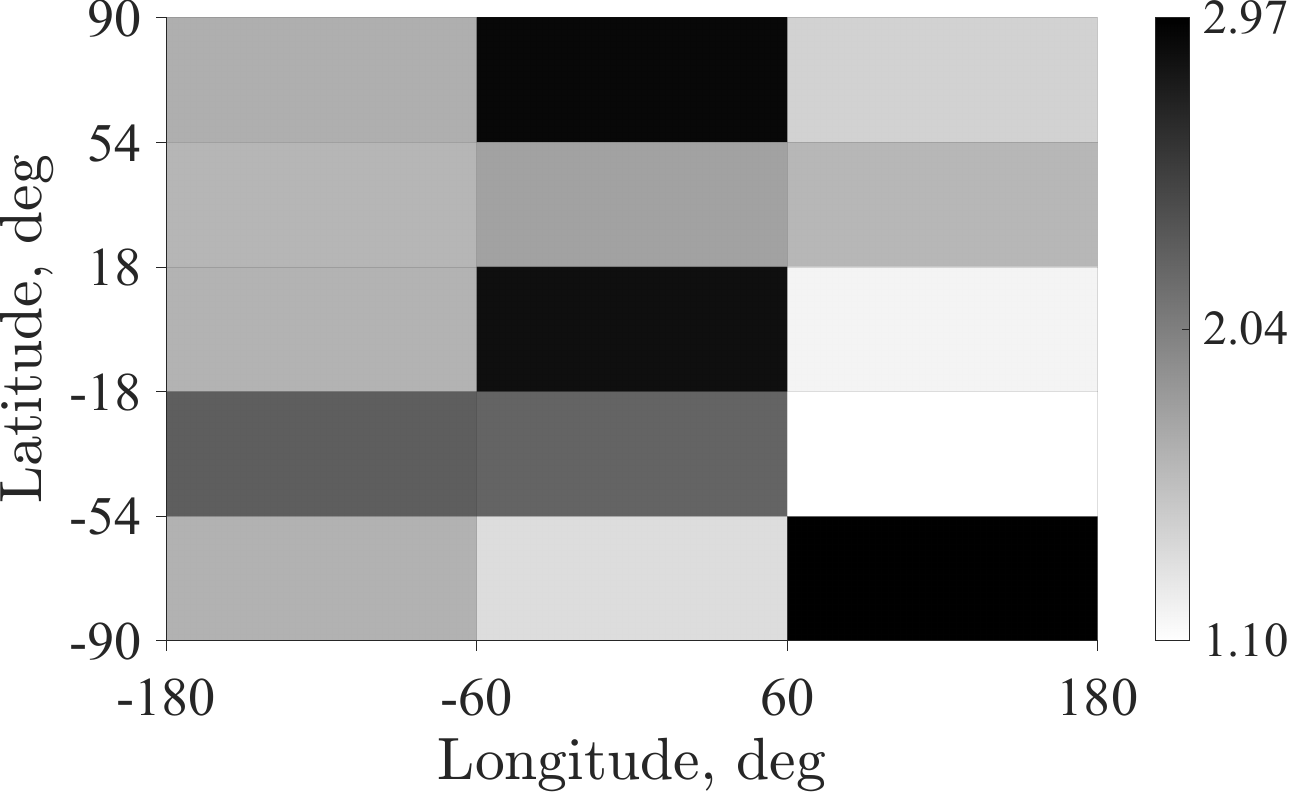}
        \caption{Instance 28.}
        \label{fig:benchmark_loc_heatmap_15}
    \end{subfigure}
    \begin{subfigure}[c]{0.44\linewidth}
        \centering
        \includegraphics[width=\linewidth]{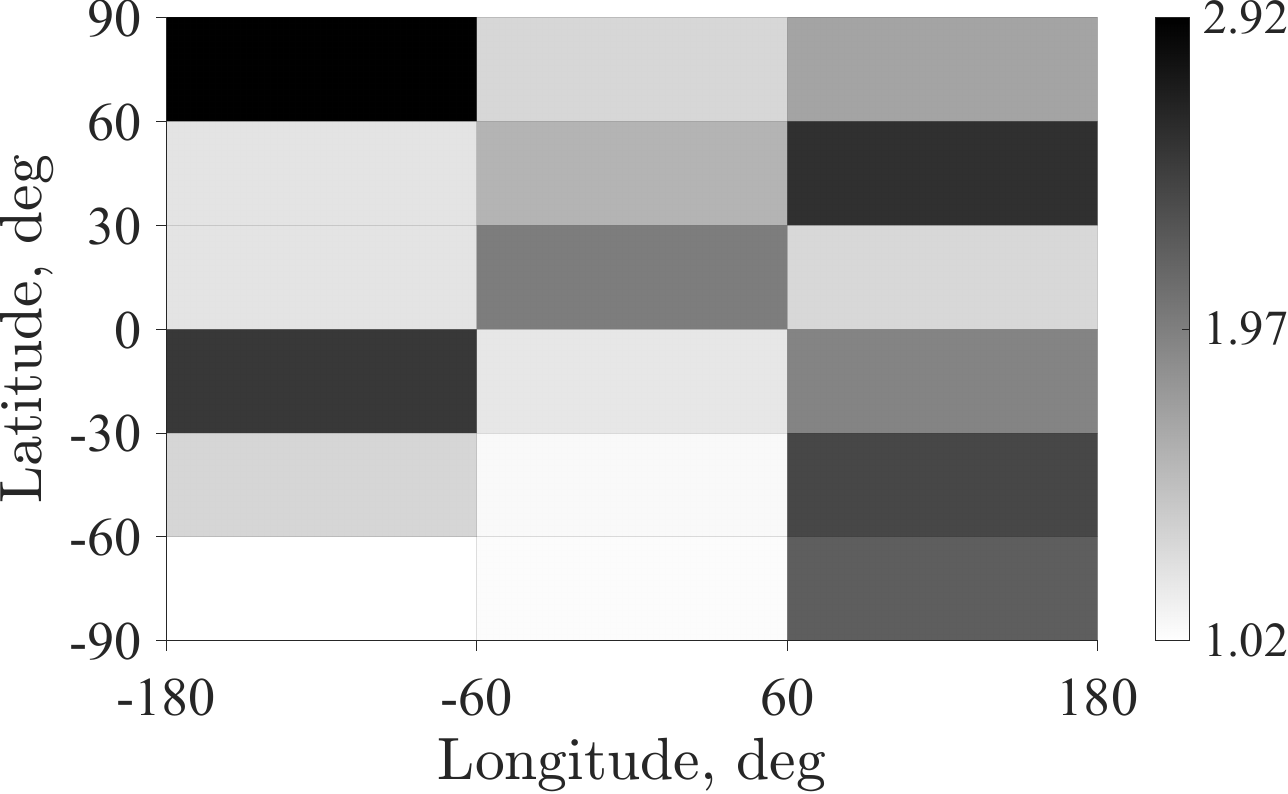}
        \caption{Instance 32.}
        \label{fig:benchmark_loc_heatmap_18}
    \end{subfigure}
    \caption{Instances' random RO measurement reward heatmaps at the epoch.}
    \label{fig:benchmark_loc_heatmap}
\end{figure}

The 32 instances are generated from all possible combinations between 100 and 250 time steps, five- and six-spacecraft constellations, \num{216} and \num{360} orbital slots, 15 and 18 uniform atmospheric regions, and uniform and random RO measurement reward definitions for each region setting (see Appendix A for a tabulated presentation of each instance's parameters). While the uniform RO measurement reward is set to one for all corresponding instances, the random RO measurement reward is computed by defining a lower and upper bound of one and three, respectively. As an illustration, Fig.~\ref{fig:benchmark_loc_heatmap} presents the magnitudes of the RO measurement rewards at the epoch assigned to instances 28 and 32. Note that these values also change at each time step.

We adopt the following metrics to benchmark the obtained results. First, the \textit{relative lower bound} (RLB), defined as $(\text{LB}-\text{LB}_{\text{B}})/\text{LB}_{\text{B}}$, where LB encodes the obtained B\&BC's lower bound, and $\text{LB}_{\text{B}}$ denotes the obtained baseline's lower bound. Note that both lower bounds (\textit{i.e.,} $\text{LB}_{\text{B}}$ and $\text{LB}$) are equal to the objective value. Second, the \textit{relative upper bound} (RUB), defined as $(\text{UB}-\text{UB}_{\text{B}})/\text{UB}_{\text{B}}$, where UB and $\text{UB}_{\text{B}}$ denote the upper bounds obtained by B\&BC and the baseline, respectively. Third, the \textit{relative runtime} (RRT), defined as $(\text{RT}-\text{RT}_{\text{B}})/\text{RT}_{\text{B}}$, where RT and $\text{RT}_{\text{B}}$ represent the solver's runtime for B\&BC and the baseline, respectively. Note that a negative value of RUB signifies a tighter upper bound (\textit{i.e.,} a better bound). Similarly, a negative RRT indicates a reduction in the runtime required to obtain an optimal solution.

\subsubsection{Results}
Table~\ref{tab:instance_results_cd} details the results of the 32 instances, where B\&BC outperforms the baseline in 27 of them in at least one benchmark metric. In particular, B\&BC obtains a higher lower bound (\textit{i.e.,} objective value) in 11 instances, a tighter upper bound in 13 instances, and a lower runtime in 14 instances. In addition, in 10 of the 27 instances, B\&BC retrieves the optimal solution, whereas the baseline triggers the runtime limit.

\begin{figure}[htpb]
    \centering
    \includegraphics[width=0.46\linewidth]{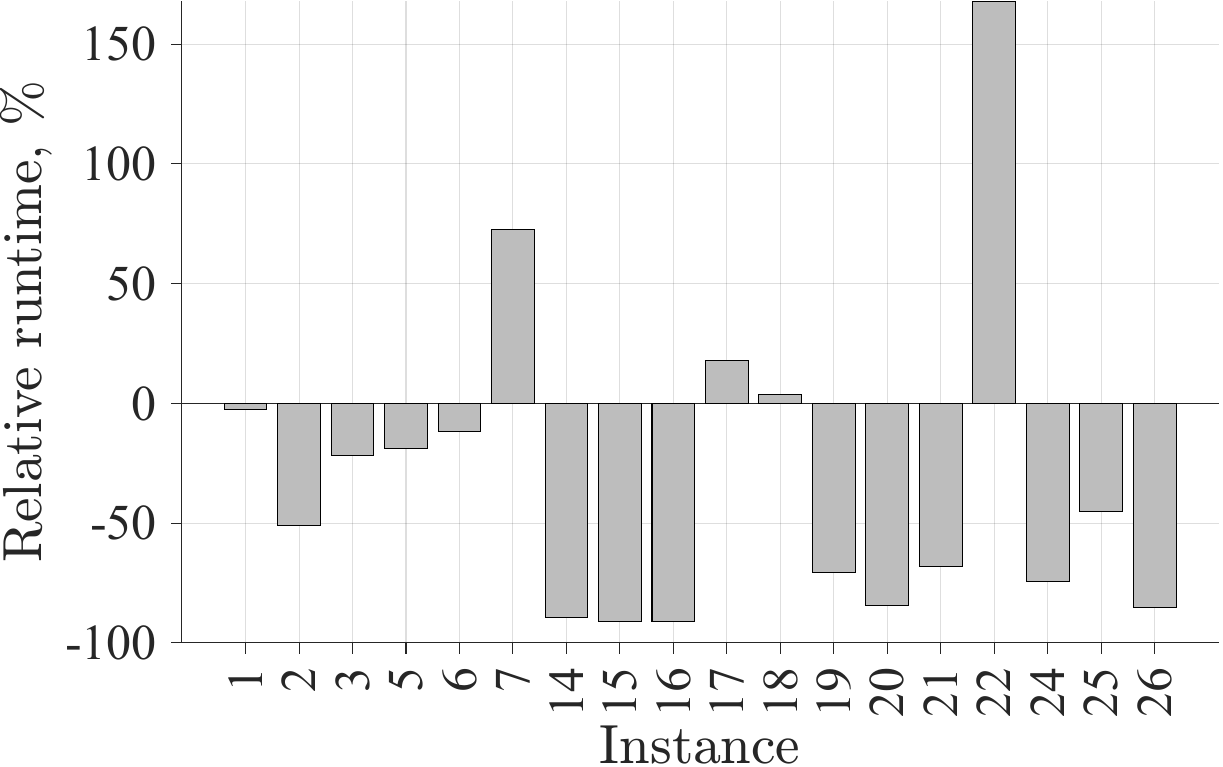}
    \caption{Relative runtime benchmark metric for RO-CDP instances.}
    \label{fig:benchmark_loc_runtime}
\end{figure}

Figure~\ref{fig:benchmark_loc_runtime} illustrates the RRT for the 18 instances where both solution approaches converge to optimality; of the total, B\&BC outperforms the baseline in 14 of them. Focusing on the B\&BC improvements, the smallest and largest percentage differences are \SI{-2.6}{\%} and \SI{-91.31}{\%}, corresponding to instances 1 and 15, respectively, both of which share a uniform reward generation and number of time steps, but differ in the remaining parameters. Furthermore, 9 instances show improvements in computational runtime of at least \SI{51.2}{\%}. Among these, six have 250 time steps, five adopt six spacecraft, random reward generation, and 18 atmospheric regions, and four share a total of 360 orbital slots.

\begin{figure}[htpb]
    \centering
    \includegraphics[width=0.46\linewidth]{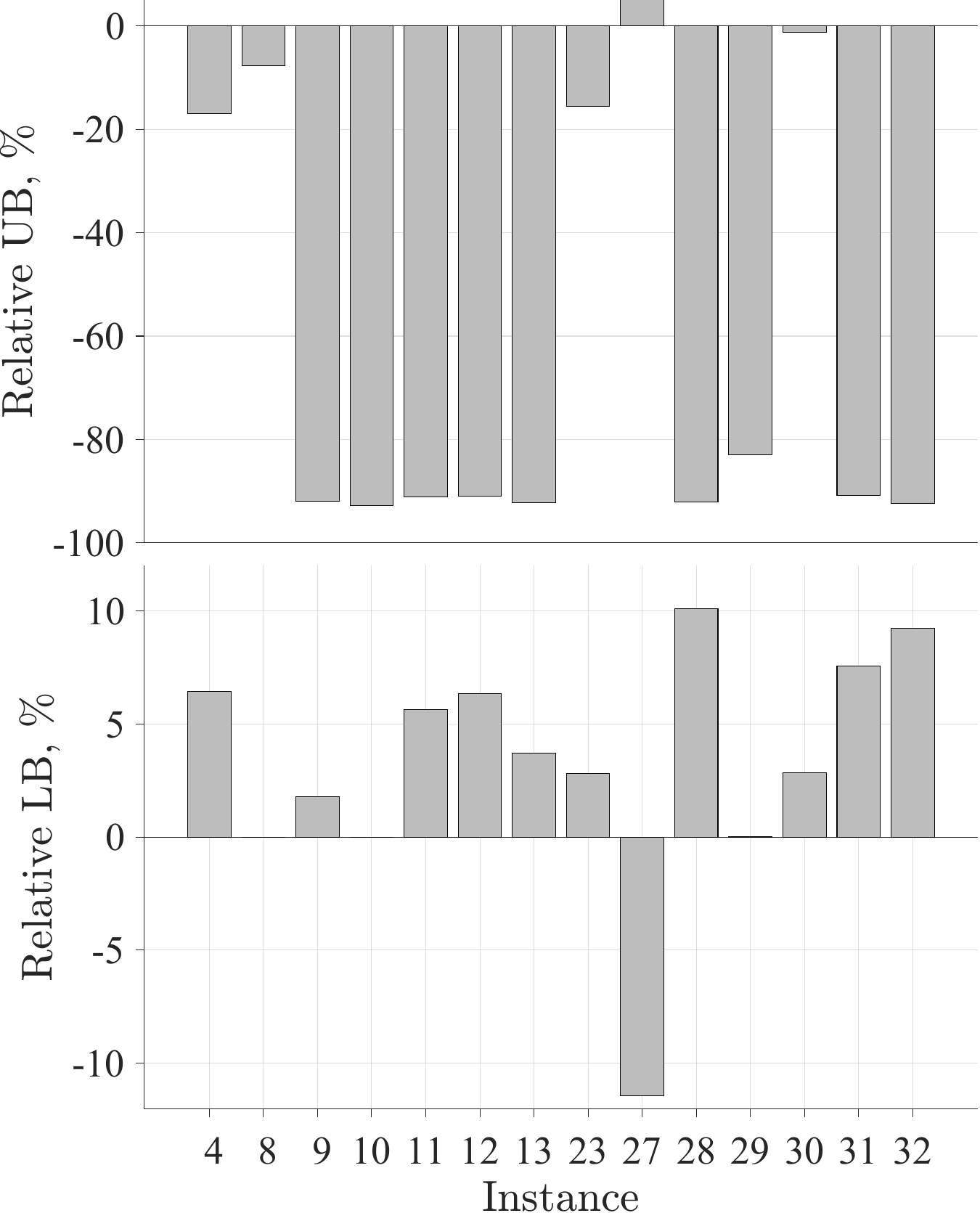}
    \caption{Relative bound benchmark metrics for RO-CDP instances.}
    \label{fig:benchmark_loc_bounds}
\end{figure}

Figure~\ref{fig:benchmark_loc_bounds} presents the RLB and RUB for the 14 instances where at least one solution approach triggers the runtime limit; of the total, B\&BC outperforms the baseline in at least one benchmark metric in 13 of them. In the context of the RLB, the smallest and largest percentage increases are \SI{0.03}{\%} and \SI{10.11}{\%}, corresponding to instances 29 and 28, respectively, where both have 360 orbital slots and a random reward generation. Similarly, with respect to the RUB, the smallest and largest percentage differences are \SI{-1.25}{\%} and \SI{-92.83}{\%}, corresponding to instances 30 and 10, respectively. For the latter benchmark metric, the two instances share the same number of time steps, the number of spacecraft, and the number of orbital slots. These results indicate that B\&BC obtains better solution bounds in most of the instances, regardless of whether both approaches obtain optimal solutions or if at least one of them is triggered by the runtime limit.

In summary, the obtained results outline the benefits of adopting B\&BC to solve RO-CDP.
First, a significant reduction in the runtime required to obtain optimal solutions enables mission designers to undergo numerous iterations during the mission design phase, unveiling architectures that achieve higher scientific data throughput at lower costs. Second, B\&BC permits mission designers to optimize for very large mission scenarios and obtain high-quality solutions, which are otherwise constrained when the baseline approach is implemented. Lastly, although the baseline outperforms B\&BC in five instances, this small number of cases does not outweigh the benefits of B\&BC.

\begin{landscape}
\begin{table}[htpb]
    \fontsize{9}{10}\selectfont
    \renewcommand{\arraystretch}{1.2}
    \caption{RO-CDP numerical experiment results.}
    \centering
    \label{tab:instance_results_cd}
    \begin{threeparttable}
    \centering
        \begin{tabular}{crrrrrrrrrrr}
        \hline
        \hline
        \multirow{2}{*}{Instance} & \multicolumn{4}{c}{Baseline}     & \multicolumn{4}{c}{B\&BC}  & \multicolumn{3}{c}{Benchmark\tnote{*}} \\
        \cmidrule(lr){2-5} \cmidrule(lr){6-9} \cmidrule(lr){10-12}
         & LB\tnote{\textdagger} & UB & MIP gap, \% & Runtime\tnote{\textdaggerdbl}, s & LB\tnote{\textdagger} & UB & MIP gap, \% & Runtime\tnote{\textdaggerdbl}, s & RLB, \% & RUB, \% & RRT\tnote{\S}, \% \\ 
        \hline
        \rowcolor{gray_row}1  & 45     & 45      & 0       & 3450.15  & 45     & 45      & 0       & 3360.41  & 0      & 0      & \bf{-2.60}   \\
        \rowcolor{gray_row}2  & 101    & 101     & 0       & 8005.92  & 101    & 101     & 0       & 3906.59  & 0      & 0      & \bf{-51.20}  \\
        \rowcolor{gray_row}3  & 60     & 60      & 0       & 9756.98  & 60     & 60      & 0       & 7619.95  & 0      & 0      & \bf{-21.90}  \\
        \rowcolor{gray_row}4  & 124    & 159     & 28.22   & -       & 132    & 132     & 0       & 4009.68  & \bf{6.45}    & \bf{-16.98}  & $\sim$ \\
        \rowcolor{gray_row}5  & 45     & 45      & 0       & 5236.46  & 45     & 45      & 0       & 4250.59  & 0      & 0      & \bf{-18.83} \\
        \rowcolor{gray_row}6  & 101    & 101     & 0       & 8275.46  & 101    & 101     & 0       & 7292.76  & 0      & 0      & \bf{-11.87} \\
        7  & 59     & 59      & 0       & 6034.49  & 59     & 59      & 0       & 10423.77 & 0      & 0      & 72.74  \\
        \rowcolor{gray_row}8 & 133    & 144     & 8.27    & -       & 133    & 133     & 0       & 4321.11  & 0       & \bf{-7.64}   & $\sim$ \\
        \rowcolor{gray_row}9  & 56     & 712     & 1171.42 & -       & 57     & 57      & 0       & 8583.33  & \bf{1.79}    & \bf{-91.99}  & $\sim$\\
        \rowcolor{gray_row}10 & 130    & 1812    & 1293.84 & -       & 130    & 130     & 0       & 6353.96  & 0       & \bf{-92.83}  & $\sim$\\
        \rowcolor{gray_row}11 & 71     & 846     & 1091.54 & -       & 75     & 75      & 0       & 9871.97  & \bf{5.63}    & \bf{-91.13}  & $\sim$ \\
        \rowcolor{gray_row}12 & 158    & 2128    & 1246.83 & -       & 168    & 193.51  & 15.18   & -       & \bf{6.33}    & \bf{-90.91}  & 0   \\
        \rowcolor{gray_row}13 & 54     & 716     & 1225.92 & -       & 56     & 56      & 0       & 1779.90   & \bf{3.70}     & \bf{-92.18}  & $\sim$\\
        \rowcolor{gray_row}14 & 128    & 128     & 0       & 5200.13  & 128    & 128     & 0       & 547.54   & 0      & 0      & \bf{-89.47} \\
        \rowcolor{gray_row}15 & 76     & 76      & 0       & 12130.64 & 76     & 76      & 0       & 1054.20   & 0      & 0      & \bf{-91.31} \\
        \rowcolor{gray_row}16 & 168    & 168     & 0       & 9881.76  & 168    & 168     & 0       & 874.57   & 0      & 0      & \bf{-91.15} \\
        17 & 95.08  & 95.08   & 0       & 262.40    & 95.08  & 95.08   & 0       & 309.36   & 0      & 0      & 17.90   \\
        18 & 205.12 & 205.12  & 0       & 571.06   & 205.12 & 205.12  & 0       & 591.27   & 0      & 0      & 3.54   \\
        \rowcolor{gray_row}19 & 125.58 & 125.58  & 0       & 8460.18  & 125.58 & 125.58  & 0       & 2467.09  & 0      & 0      & \bf{-70.84} \\
        \rowcolor{gray_row}20 & 262.29 & 262.29  & 0       & 7539.99  & 262.29 & 262.29  & 0       & 1183.03  & 0      & 0      & \bf{-84.31} \\
        \rowcolor{gray_row}21 & 91.34  & 91.34   & 0       & 3827.64  & 91.34  & 91.34   & 0       & 1214.55  & 0      & 0      & \bf{-68.27} \\
        22 & 200.37 & 200.37  & 0       & 2834.38  & 200.37 & 200.37  & 0       & 7597.73  & 0      & 0      & 168.06 \\
        \rowcolor{gray_row}23 & 120.95 & 147.33  & 21.81   & -       & 124.35 & 124.35  & 0       & 6585.89  & \bf{2.81}    & \bf{-15.60}   & $\sim$ \\
        \rowcolor{gray_row}24 & 274.68 & 274.68  & 0       & 1122.78  & 274.68 & 274.68  & 0       & 286.08   & 0      & 0      & \bf{-74.52} \\
        \rowcolor{gray_row}25 & 124.52 & 124.52  & 0       & 2569.75  & 124.52 & 124.52  & 0       & 1410.25  & 0      & 0      & \bf{-45.12} \\
        \rowcolor{gray_row}26 & 260.23 & 260.23  & 0       & 5018.91  & 260.23 & 260.23  & 0       & 746.79   & 0      & 0      & \bf{-85.12} \\
        27 & 158.09 & 168.83  & 6.79    & -       & 139.99 & 1398.78 & 899.16  & -       & -11.45  & 728.51  & 0     \\
        \rowcolor{gray_row}28 & 308.33 & 4285.34 & 1289.84 & -       & 339.49 & 339.49  & 0       & 8177     & \bf{10.11}   & \bf{-92.08}  & $\sim$     \\
        \rowcolor{gray_row}29 & 105    & 1441.55 & 1272.90  & -       & 105.03 & 244.75  & 133.03  & -       & \bf{0.03}    & \bf{-83.02}  & 0     \\
        \rowcolor{gray_row}30 & 221.75 & 3690.74 & 1564.37 & -       & 228.07 & 3644.47 & 1497.93 & -       & \bf{2.85}    & \bf{-1.25}   & 0     \\
        \rowcolor{gray_row}31 & 148.94 & 1733.82 & 1064.06 & -       & 160.20  & 160.20   & 0       & 8486.45  & \bf{7.56}    & \bf{-90.76}  & $\sim$     \\
        \rowcolor{gray_row}32 & 305.49 & 4344.64 & 1322.18 & -       & 333.65 & 333.65      & 0      & 1252.64       & \bf{9.22} & \bf{-92.32} & $\sim$  \\ 
        \hline
        \hline
    \end{tabular}
    \begin{tablenotes}
        \item[*] Boldface indicates that B\&BC outperforms the baseline.
        \item[\textdagger] Objective value.
        \item[\textdaggerdbl] A hyphen (-) indicates that the solver's runtime limit of \SI{14400}{s} is triggered.
        \item[\S] A tilde ($\sim$) indicates that B\&BC finds an optimal solution while the baseline is halted as it reaches the runtime limit.
    \end{tablenotes}
    \end{threeparttable}
\end{table}
\end{landscape}

\clearpage
\subsection{Branch-and-Benders-Cut RO Constellation Reconfiguration Benchmark}\label{sec:numexp_benders_cr}

This benchmark study conducts a sensitivity analysis between the $\Delta v$ transfer budget and the collected RO measurement rewards, and compares RO-CRP's solution quality using B\&BC and the Gurobi Optimizer to solve the baseline model (\textit{i.e.,} Sec.~\ref{sec:methodology_crp}).

\subsubsection{Setup}
This benchmark study employs all default settings of the Gurobi Optimizer, with two exceptions. First, the runtime limit is set to one hour. The rationale for reducing the runtime limit relative to the design benchmark (\textit{i.e.,} Sec.~\ref{sec:numexp_benders_cd}) is that we now run 180 experiments in total. The second exception is to enable lazy constraints to implement B\&BC with the Gurobi Optimizer.

The 18 instances are generated by combining 250 and 800 time steps, 1125, 1500, and 1875 orbital slots, yielding \num{632250}, \num{1124250}, and \num{1756875} spacecraft pairs, respectively, and 15, 30, and 45 atmospheric regions randomly selected from a pool of 48 (see Appendix A for an outline of each instance's parameters). Each individual instance is solved by varying the $\Delta v$ transfer budget $\varepsilon_s$ from \SI{0.20}{km/s} to \SI{1}{km/s} in steps of \SI{0.20}{km/s}, yielding five cases per instance.

Without loss of generality, all instances adopt a common three-spacecraft constellation configuration. For each spacecraft, the set of destination orbital slots is generated as follows. From the current orbital slot, four additional uniform altitude steps of \SI{150}{km} are defined. For each altitude step, a positive and negative $\Delta$ inclination and a $\Delta$ right ascension of the ascending node (RAAN) are applied, assuming a $\Delta v$ transfer budget of \SI{1}{km/s} and a scaling factor of 0.75, producing four additional orbital planes (we refer the reader to Appendix B for the computation of the $\Delta$ inclination and RAAN). Subsequently, the 15, 20, and 25 phasing steps are defined for each orbital plane. Then, the transfer cost is obtained by sequentially performing a Hohmann transfer maneuver, plane change, and phasing. Lastly, the set of destination orbital slots is constructed as the union of the spacecraft's transfer slots. 

We adopt the RLB, RUB, and RRT metrics defined in Sec.~\ref{sec:numexp_benders_cd} to benchmark the obtained results. 

\subsubsection{Results}

\begin{figure}[htpb]
    \centering
    \includegraphics[width=0.7\linewidth]{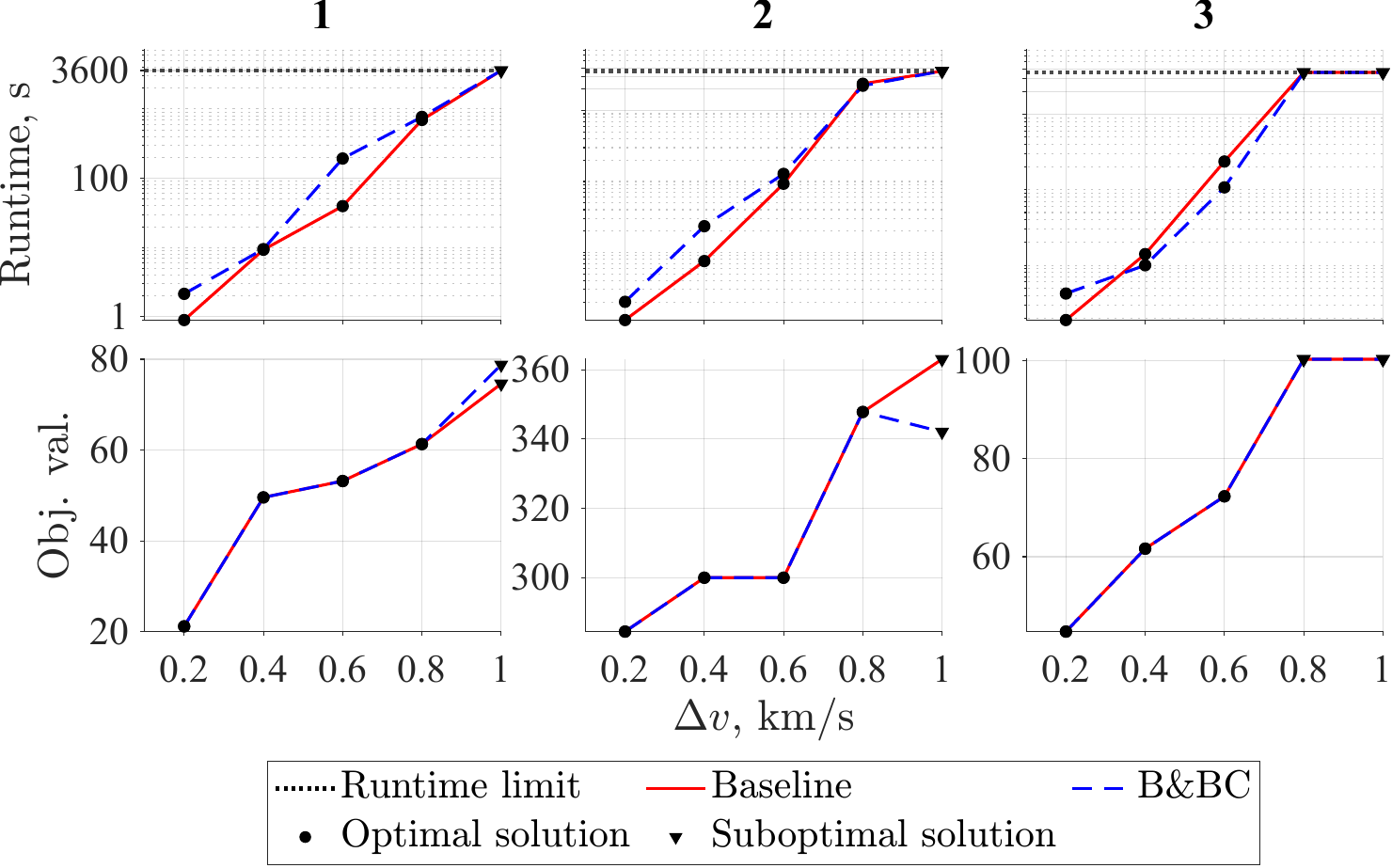}
    \caption{Runtime and objective values for instances 1 to 3.}
    \label{fig:benchmark_rec_1_3}
\end{figure}

\begin{figure}[htpb]
    \centering
    \includegraphics[width=0.7\linewidth]{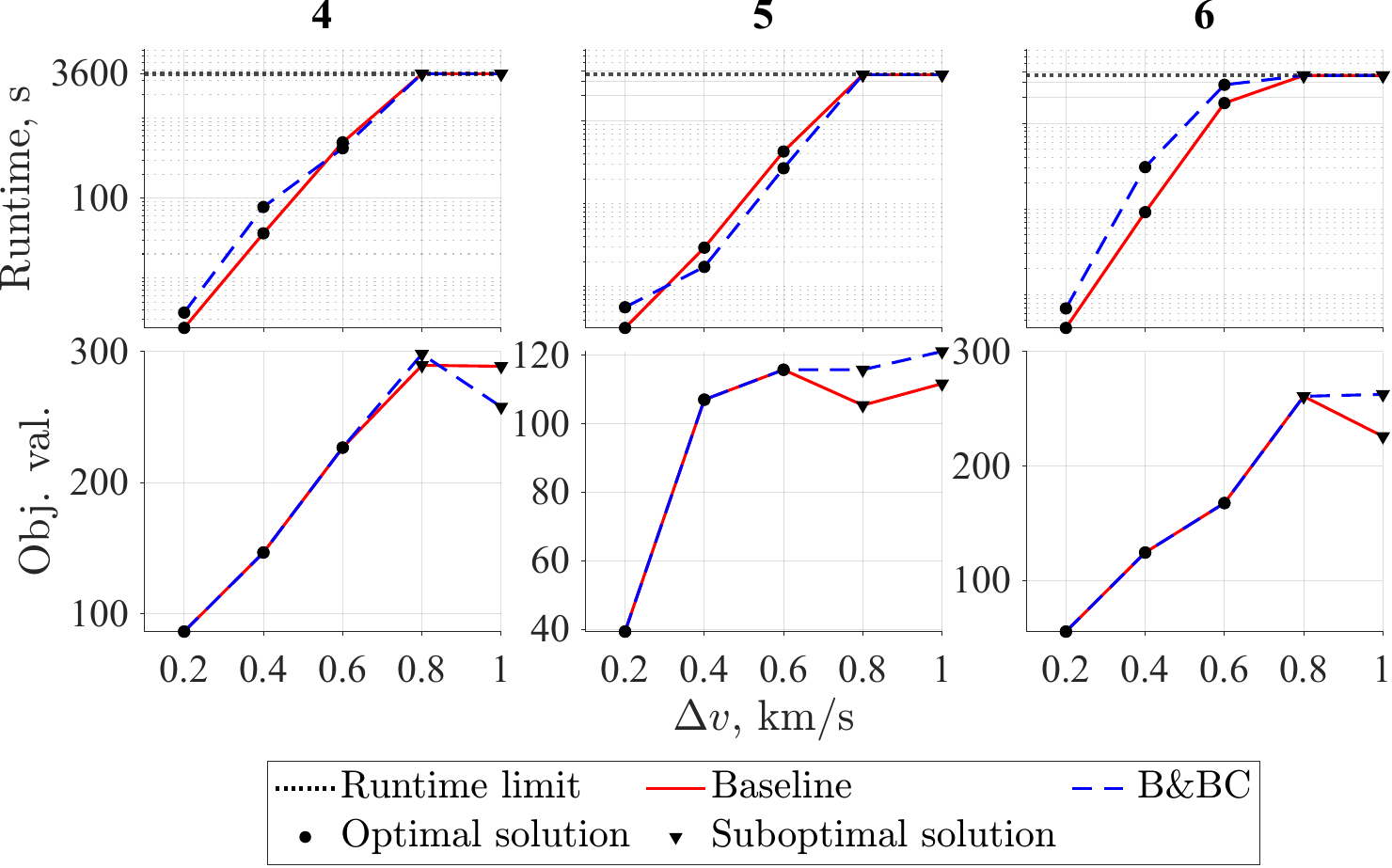}
    \caption{Runtime and objective values for instances 4 to 6.}
    \label{fig:benchmark_rec_4_6}
\end{figure}

\begin{figure}[htpb]
    \centering
    \includegraphics[width=0.7\linewidth]{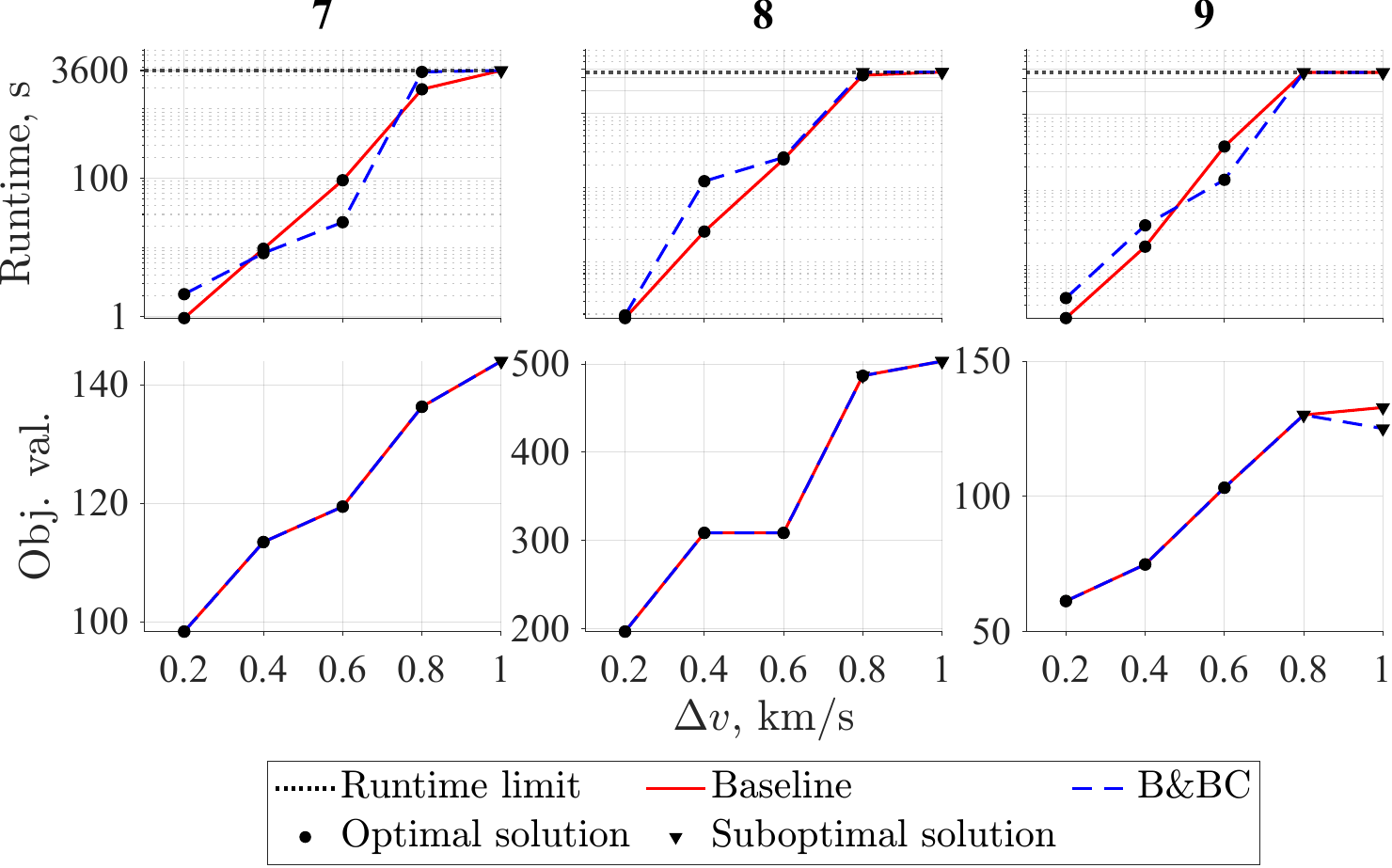}
    \caption{Runtime and objective values for instances 7 to 9.}
    \label{fig:benchmark_rec_7_9}
\end{figure}

\begin{figure}[htpb]
    \centering
    \includegraphics[width=0.7\linewidth]{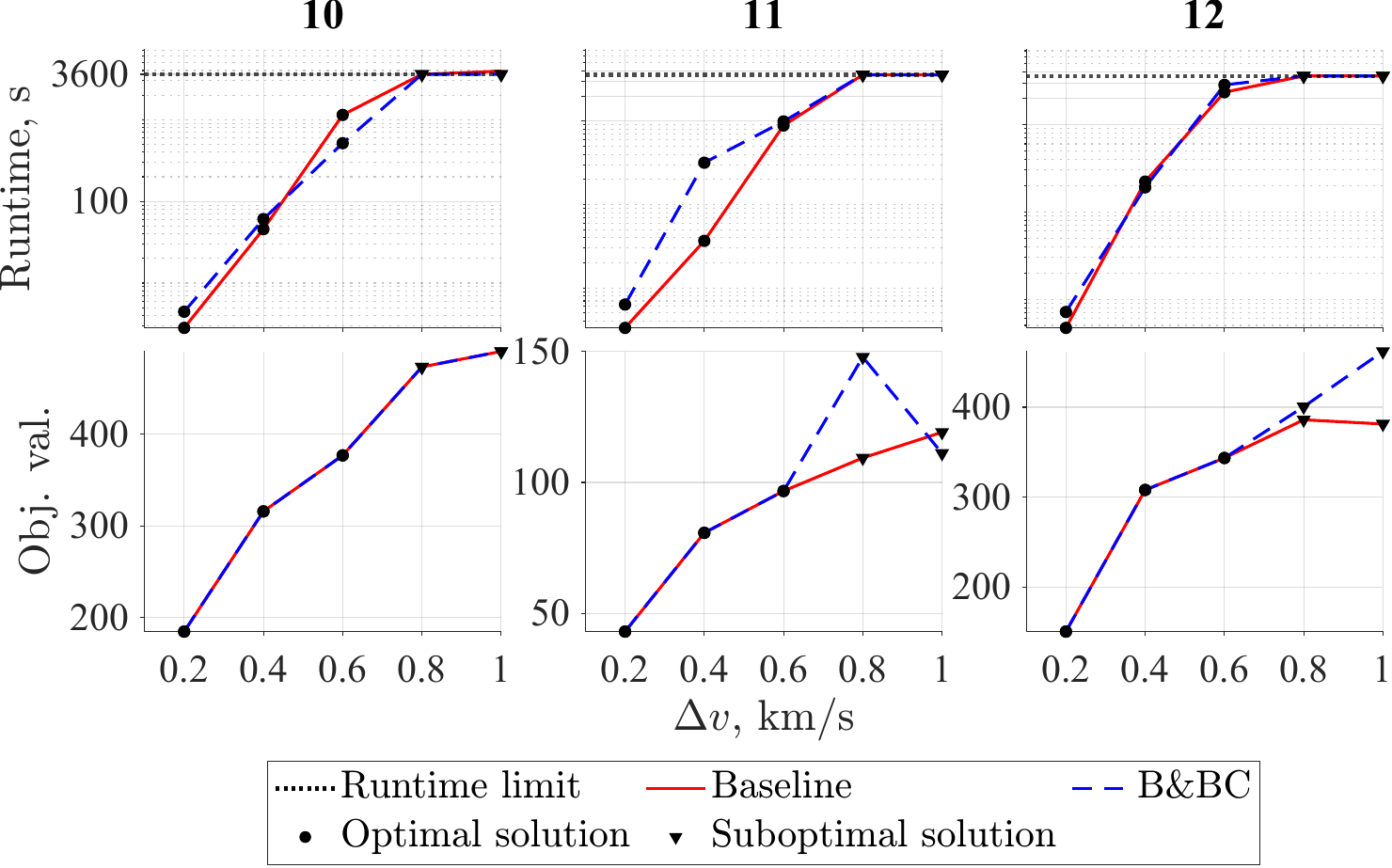}
    \caption{Runtime and objective values for instances 10 to 12.}
    \label{fig:benchmark_rec_10_12}
\end{figure}

\begin{figure}[htpb]
    \centering
    \includegraphics[width=0.7\linewidth]{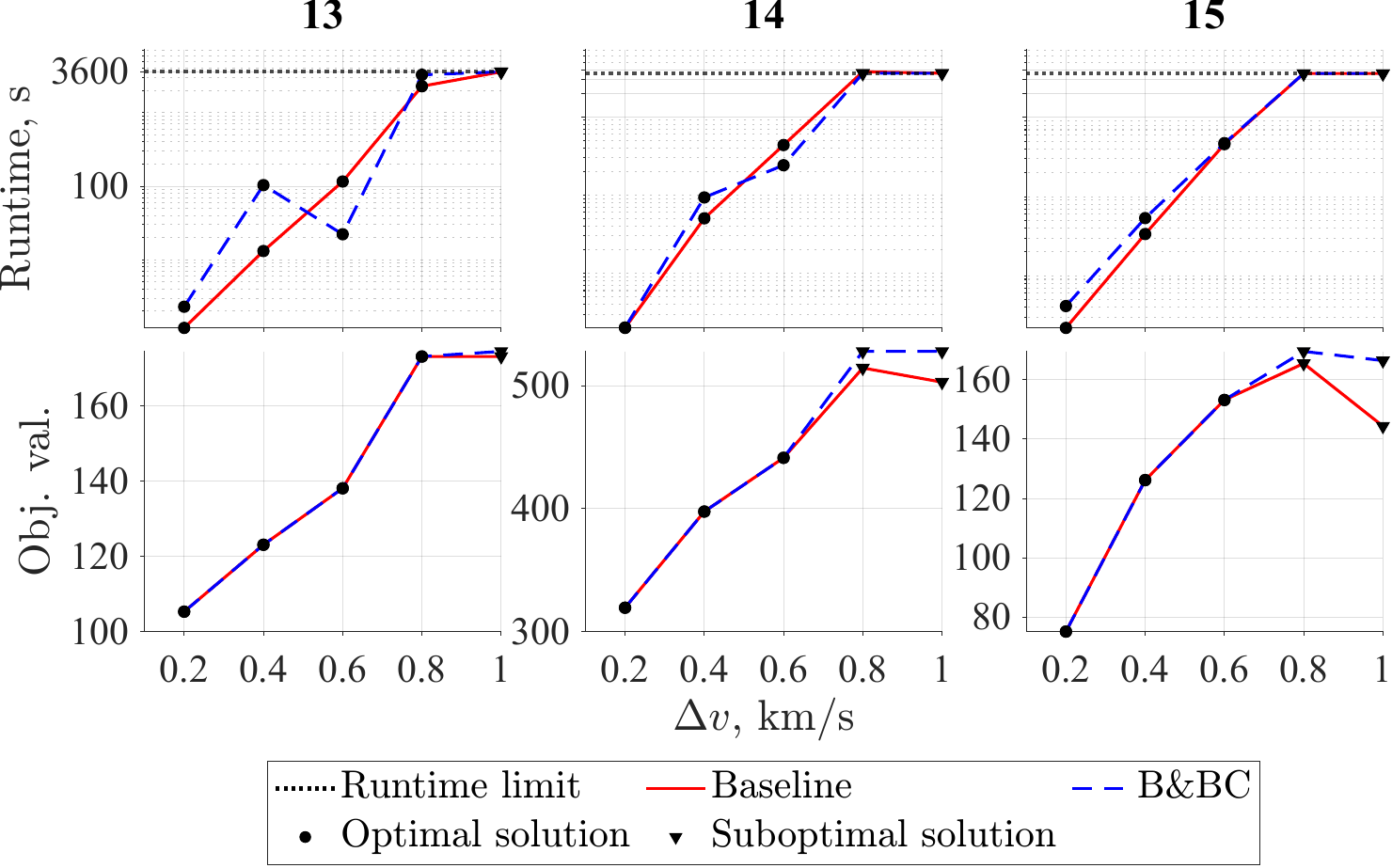}
    \caption{Runtime and objective values for instances 13 to 15.}
    \label{fig:benchmark_rec_13_15}
\end{figure}

\begin{figure}[htpb]
    \centering
    \includegraphics[width=0.7\linewidth]{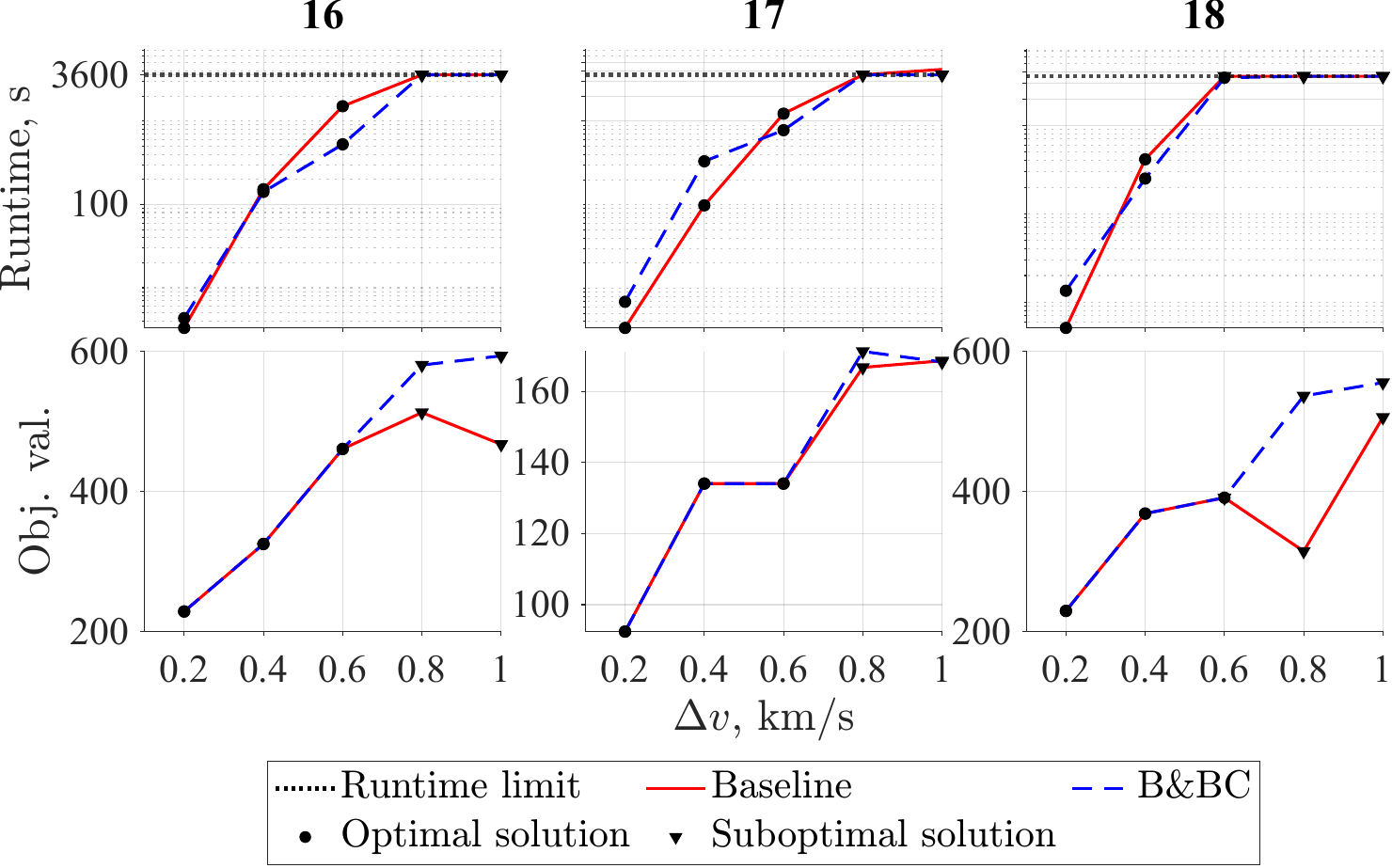}
    \caption{Runtime and objective values for instances 16 to 18.}
    \label{fig:benchmark_rec_16_18}
\end{figure}

Figures~\ref{fig:benchmark_rec_1_3} to \ref{fig:benchmark_rec_16_18} show, for each instance, the sensitivity of the runtime and objective value to the spacecraft's $\Delta v$ transfer budget $\varepsilon_s$. All instances are solved to optimality when the $\Delta v$ transfer budget $\varepsilon_s$ is at most \SI{0.40}{km/s}. For a $\Delta v$ transfer budget $\varepsilon_s$ of \SI{0.60}{km/s}, instance 18, which is the one with the largest number of constraints and decision variables, is solved to optimality by B\&BC; in contrast, the baseline triggers the runtime limit. When the $\Delta v$ transfer budget $\varepsilon_s$ is set to \SI{0.80}{km/s}, instances 1, 2, 7 and 13 are solved to optimality by both, and instance 8 is only solved to optimality by the baseline. For the largest $\Delta v$ transfer budget $\varepsilon_s$, \SI{1}{km/s}, all instances trigger the runtime limit. These results show that, although each instance is different from one another in the overall number of constraints and decision variables, a larger $\Delta v$ transfer budget $\varepsilon_s$ unlocks more feasible spacecraft orbital transfers, leading to a larger set of candidate constellation reconfigurations, which ultimately enlarge the solution space and make the problem more challenging regardless of the chosen method.

Focusing on the plot depicting the spacecraft's $\Delta v$ transfer budget $\varepsilon_s$ and the objective value, there is a general trend among the instances' optimal solutions that a larger $\Delta v$ transfer budget $\varepsilon_s$ leads to improvements in the objective value. Although true, there are three outliers (\textit{i.e.,} instances 2, 8, and 17) where increasing the $\Delta v$ transfer budget $\varepsilon_s$ from \SI{0.40}{km/s} to \SI{0.60}{km/s} yields the same optimal objective value. Given the random definition of their optimization parameters (\textit{i.e.,} number of time steps $T$, orbital slots $J$, and atmospheric regions $O$), an increment of \SI{0.20}{km/s} in their $\Delta v$ transfer budget $\varepsilon_s$ does not unlock new feasible orbital slots capable of increasing the optimal objective value.

Table~\ref{tab:instance_results_cr_rlb} presents the RLB for each instance; note that a value of zero indicates that both B\&BC and the baseline obtain the same lower bound (\textit{i.e.,} objective value). For the cases where the RLB is nonzero, B\&BC obtains a better RLB for at least one $\Delta v$ transfer budget $\varepsilon_s$ in 12 instances, where the best obtained RLB corresponds to instance 18 with a value of \SI{70.44}{\%}. The benefits of using B\&BC are outlined for the $\Delta v$ transfer budget $\varepsilon_s$ of \SI{0.80}{km/s}, where it outperforms the baseline in all instances that have nonzero RLBs. Furthermore, for a $\Delta v$ transfer budget $\varepsilon_s$ of \SI{1}{km/s}, B\&BC outperforms the baseline in nine out of 14 instances with nonzero RLBs. 
\begin{table}[htpb]
    \fontsize{9}{10}\selectfont
    \renewcommand{\arraystretch}{1.2}
    \caption{RLB for RO-CRP numerical experiment results.}
    \centering
    \label{tab:instance_results_cr_rlb}
    \begin{threeparttable}
    \centering
        \begin{tabular}{crrrrr} %
        \hline
        \hline
        \multirow{2}{*}{Instance} & \multicolumn{5}{c}{$\varepsilon_s,$ \si{km/s}} \\
        \cmidrule(lr){2-6} 
        & 0.20 & 0.40 &0.60 &0.80 & 1\\
        \hline
        \rowcolor{gray_row}1  & 0  & 0  & 0 & 0     & \bf{5.62}   \\
        2  & 0  & 0  & 0  & 0     & -5.77  \\
        3  & 0  & 0 & 0  & 0    & 0      \\
        \rowcolor{gray_row}4  & 0 & 0  & 0 & \bf{2.96}  & -10.73 \\
        \rowcolor{gray_row}5  & 0  & 0  & 0  & \bf{9.83}  & \bf{8.43}   \\
        \rowcolor{gray_row}6  & 0 & 0  & 0  & 0    & \bf{16.26}  \\
        7  & 0  & 0  & 0  & 0     & 0      \\
        8  & 0  & 0 & 0 & 0     & 0      \\
        9  & 0  & 0  & 0  & 0     & -5.86  \\
        10 & 0 & 0  & 0 & 0     & 0     \\
        \rowcolor{gray_row}11 & 0  & 0  & 0 & \bf{35.18} & -6.74  \\
        \rowcolor{gray_row}12 & 0  & 0 & 0  & \bf{3.76}  & \bf{21.16}  \\
        \rowcolor{gray_row}13 & 0  & 0  & 0 & 0     & \bf{0.78}   \\
        \rowcolor{gray_row}14 & 0  & 0  & 0  & \bf{2.62}  & \bf{4.97}   \\
        \rowcolor{gray_row}15 & 0  & 0  & 0  & \bf{2.45}  & \bf{15.29}  \\
        \rowcolor{gray_row}16 & 0 & 0  & 0 & \bf{13.22} & \bf{27.04}  \\
        \rowcolor{gray_row}17 & 0  & 0  & 0  & \bf{2.70}   & -0.21  \\
        \rowcolor{gray_row}18 & 0  & 0 & 0  & \bf{70.44} & \bf{9.80}  \\
        \hline
        \hline
    \end{tabular}
    \end{threeparttable}
\end{table}

Table~\ref{tab:instance_results_cr_rub} introduces the RUB for all instances. For this particular metric, B\&BC obtains tighter bounds across all instances for at least one $\Delta v$ transfer budget $\varepsilon_s$ value, where the best RUB corresponds to instance 18 and has a magnitude of \SI{-88.20}{\%}. Although the baseline outperforms B\&BC for a $\Delta v$ transfer budget $\varepsilon_s$ of \SI{0.80}{km/s}, delivering tighter bounds in 8 of 14 instances with nonzero RUB, B\&BC significantly outperforms the baseline in 17 of the 18 instances using \SI{1}{km/s}.
\begin{table}[htpb]
    \fontsize{9}{10}\selectfont
    \renewcommand{\arraystretch}{1.2}
    \caption{RUB for RO-CRP numerical experiment results.}
    \centering
    \label{tab:instance_results_cr_rub}
    \begin{threeparttable}
    \centering
        \begin{tabular}{crrrrr} %
        \hline
        \hline
        \multirow{2}{*}{Instance} & \multicolumn{5}{c}{$\varepsilon_s,$ \si{km/s}} \\
        \cmidrule(lr){2-6} 
        & 0.20 & 0.40 &0.60 &0.80 & 1\\
        \hline
        \rowcolor{gray_row}1  & 0  & 0  & 0    & 0      & \bf{-14.82} \\
        \rowcolor{gray_row}2  & 0  & 0  & 0     & 0      & \bf{-0.41}  \\
        \rowcolor{gray_row}3  & 0  & 0 & 0     & 58.12  & \bf{-1.80}   \\
        \rowcolor{gray_row}4  & 0 & 0  & 0    & 2.47   & \bf{-2.80}   \\
        \rowcolor{gray_row}5  & 0  & 0  & 0     & \bf{-4.72}  & \bf{-6.53}  \\
        \rowcolor{gray_row}6  & 0 & 0  & 0     & 1.63   & \bf{-5.09}  \\
        \rowcolor{gray_row}7  & 0  & 0  & 0     & 0      & \bf{-0.14}  \\
        \rowcolor{gray_row}8  & 0  & 0 & 0    & 328.37 & \bf{-2.06}  \\
        \rowcolor{gray_row}9  & 0  & 0  & 0 & 72.52  & \bf{-2.70}   \\
        \rowcolor{gray_row}10 & 0 & 0  & 0 & \bf{-3.44}  & 1.91   \\
        \rowcolor{gray_row}11 & 0  & 0  & 0    & \bf{-0.36}  & \bf{-7.33}  \\
        \rowcolor{gray_row}12 & 0  & 0 & 0     & \bf{-0.37}  & \bf{-9.04}  \\
        \rowcolor{gray_row}13 & 0  & 0  & 0    & 0      & \bf{-0.18}  \\
        \rowcolor{gray_row}14 & 0  & 0  & 0     & 34.36  & \bf{-4.13}  \\
        \rowcolor{gray_row}15 & 0  & 0  & 0     & 0.67   & \bf{-2.31}  \\
        \rowcolor{gray_row}16 & 0 & 0  & 0    & 21.34  & \bf{-6.25}  \\
        \rowcolor{gray_row}17 & 0  & 0  & 0     & \bf{-0.85}  & \bf{-6.48}  \\
        \rowcolor{gray_row}18 & 0  & 0  & \bf{-88.20} & \bf{-19.49} & \bf{-2.09} \\
        \hline
        \hline
    \end{tabular}
    \end{threeparttable}
\end{table}

Table~\ref{tab:instance_results_cr_rrt} reports the RRT corresponding to each instance. B\&BC outperforms the baseline in 13 instances for at least one $\Delta v$ transfer budget $\varepsilon_s$, where it achieves the highest reduction, \SI{-80.84}{\%} for instance 13, and solves instance 18 to optimality while the baseline triggers the runtime limit. However, the baseline consistently outperforms B\&BC in all instances that use a $\Delta v$ transfer budget $\varepsilon_s$ of \SI{0.20}{km/s}. Given the low $\Delta v$ transfer budget $\varepsilon_s$, both problems' solution spaces are small; therefore, it is more efficient to explore the baseline's branch-and-bound tree rather than the master problem's branch-and-bound tree and separate optimality cuts through the B\&BC algorithm. The benefits of adopting B\&BC are outlined for a $\Delta v$ transfer budget $\varepsilon_s$ of \SI{0.60}{km/s}, where B\&BC outperforms the baseline in 10 of the 17 instances solved to optimality.
\begin{table}[htpb]
    \fontsize{9}{10}\selectfont
    \renewcommand{\arraystretch}{1.2}
    \caption{RRT for RO-CRP numerical experiment results.}
    \centering
    \label{tab:instance_results_cr_rrt}
    \begin{threeparttable}
    \centering
        \begin{tabular}{crrrrr} %
        \hline
        \hline
        \multirow{2}{*}{Instance} & \multicolumn{5}{c}{$\varepsilon_s,$ \si{km/s}} \\
        \cmidrule(lr){2-6} 
        & 0.20 & 0.40 &0.60 &0.80 & 1\\
        \hline
        1  & 139.19 & 2.17   & 387.46 & 10.53 & 0 \\
        \rowcolor{gray_row}2  & 80.78  & 210    & 37.73  & \bf{-6.22} & 0 \\
        \rowcolor{gray_row}3  & 124.67 & \bf{-28.88} & \bf{-54.57} & 0     & 0 \\
        \rowcolor{gray_row}4  & 55.47  & 113.94 & \bf{-15.40}  & 0     & 0 \\
        \rowcolor{gray_row}5  & 77.43  & \bf{-41.31} & \bf{-37.22} & 0     & 0 \\
        6  & 68.85  & 235.62 & 63.42  & 0     & 0 \\
        \rowcolor{gray_row}7  & 121.63 & \bf{-13.67} & \bf{-75.25} & 78.23 & 0 \\
        8  & 8.91   & 375.83 & 6.05   & $\dagger$  & 0 \\
        \rowcolor{gray_row}9  & 84.14  & 91.49  & \bf{-63.68} & 0     & 0 \\
        \rowcolor{gray_row}10 & 57.80   & 32.78  & \bf{-54.92} & 0     & 0 \\
        11 & 90.69  & 765.74 & 11.77  & 0     & 0 \\
        \rowcolor{gray_row}12 & 52.22  & \bf{-14.12} & 21.40   & 0     & 0 \\
        \rowcolor{gray_row}13 & 94.12  & 687.80  & \bf{-80.84} & 43    & 0 \\
        \rowcolor{gray_row}14 & 1.14   & 84.92  & \bf{-44.75} & 0     & 0 \\
        15 & 88.91  & 59.07  & 2.98   & 0     & 0 \\
        \rowcolor{gray_row}16 & 30.98  & \bf{-7.02}  & \bf{-65.16} & 0     & 0 \\
        \rowcolor{gray_row}17 & 105.43 & 237.26 & \bf{-36.57} & 0     & 0 \\
        \rowcolor{gray_row}18 & 163.21 & \bf{-39.48} & \bf{$\sim$}  & 0     & 0\\
        \hline
        \hline
    \end{tabular}
    \begin{tablenotes}
        \item[] A tilde ($\sim$) indicates that B\&BC finds an optimal solution while the baseline is halted as it reaches the runtime limit.
        \item[] A dagger ($\dagger$) indicates that the baseline finds an optimal solution while B\&BC is halted as it reaches the runtime limit.
    \end{tablenotes}
    \end{threeparttable}
\end{table}

In summary, RO-CRP proves to be a challenging combinatorial problem that can be effectively addressed using B\&BC. Overall, the utmost benefit materializes for those instances that have the largest $\Delta v$ transfer budget $\varepsilon_s$ of \SI{0.80}{km/s} and \SI{1}{km/s}. These cases mostly trigger the runtime limit due to their large solution space, and the use of B\&BC enables obtaining higher objective values, tighter upper bounds, or both. Conversely, the cases with smaller solution spaces (\textit{i.e.,} $\Delta v$ transfer budget $\varepsilon_s$ of \SI{0.20}{km/s} and \SI{0.40}{km/s}) can be efficiently solved using the baseline.

\subsection{RO Mission Design and Operation Under Dynamic Mission Objectives}\label{sec:numexp_cd_cr}
This case study addresses the mission design and operations pipeline by using RO-CDP to design a constellation configuration for global RO measurements, and RO-CRP to reconfigure the obtained constellation after regional volcanic activity is detected. 

\subsubsection{Setup: RO-CDP} \label{sec:case_study_cd_setup}
Without loss of generality, RO-CDP is used to determine a five-spacecraft constellation configuration that maximizes the RO measurements of the Venusian atmosphere. To this end, we define 12 atmospheric regions using three longitude bands, four latitude bands, and one altitude band spanning from \SI{50}{km} to \SI{90}{km}, where each region has an RO measurement reward equal to one. Note that below \SI{40}{km}, RO measurement quality decreases due to defocusing loss and absorption \cite{imamura_eps_2017}. 

The set of 450 orbital slots is generated from all combinations of the orbital elements depicted in Table~\ref{table:numexp_cd_params_oe}, where each combination corresponds to a circular orbit. Using these orbital slots, we construct the set of spacecraft pairs, totaling \num{101025}.

\begin{table}[ht]
\centering
    \caption{Orbital elements used to generate orbital slots.}
    \label{table:numexp_cd_params_oe}%
    \begin{tabular}{lc}
        \hline
        \hline
        Orbital elements & Values\\
        \hline
        Orbital radius, km & $\{6802.54, 7358.22, 7435.88\}$\\
        Inclination, deg & $\{65, 75, 85, 100, 115\}$\\
        RAAN, deg & $\{0, 72, 144, 216, 288\}$ \\
        Argument of latitude, deg & $\{10, 70, 130, 190, 250, 310\}$\\
        \hline
        \hline
    \end{tabular}
\end{table}

We assume an optimization horizon of \SI{123.70}{h} (\textit{i.e.,} approximately five Earth days) discretized with uniform time steps of size \SI{120}{s}. In particular, the optimization horizon is selected as follows:
\begin{equation}
    N_1T_1\approx N_2T_2\approx N_3T_3\approx 123.70~\text{h}
\end{equation}
where $N_1,N_2$ and $N_3$ are integer numbers and $T_1,T_2$ and $T_3$ are the orbital periods of the candidate orbital slots. Following this approach, the inter-satellite geometry approximately repeats every optimization horizon, but their relative geometry to Venus changes. Therefore, we simulate the optimal constellation configuration for \SI{5832.54}{h} (i.e., one Venusian day) to determine the location of the RO tangent points over a longer period.

\subsubsection{Results: RO-CDP}\label{sec:results_rocdp}
The optimal constellation configuration of five spacecraft collects a total RO measurement reward of \num{1590}. Throughout the optimization scenario, the optimal constellation has a total number of \num{1667} RO measurements. The difference in the two magnitudes derives from RO-CDP's formulation. Specifically, if two RO measurement rewards over a region occur at the same time step, they activate the same RO measurement decision variable that has a maximum value of one.

\begin{figure}[htpb]
    \centering
    \begin{subfigure}[c]{0.492\linewidth}
        \centering
        \includegraphics[width=0.8\linewidth]{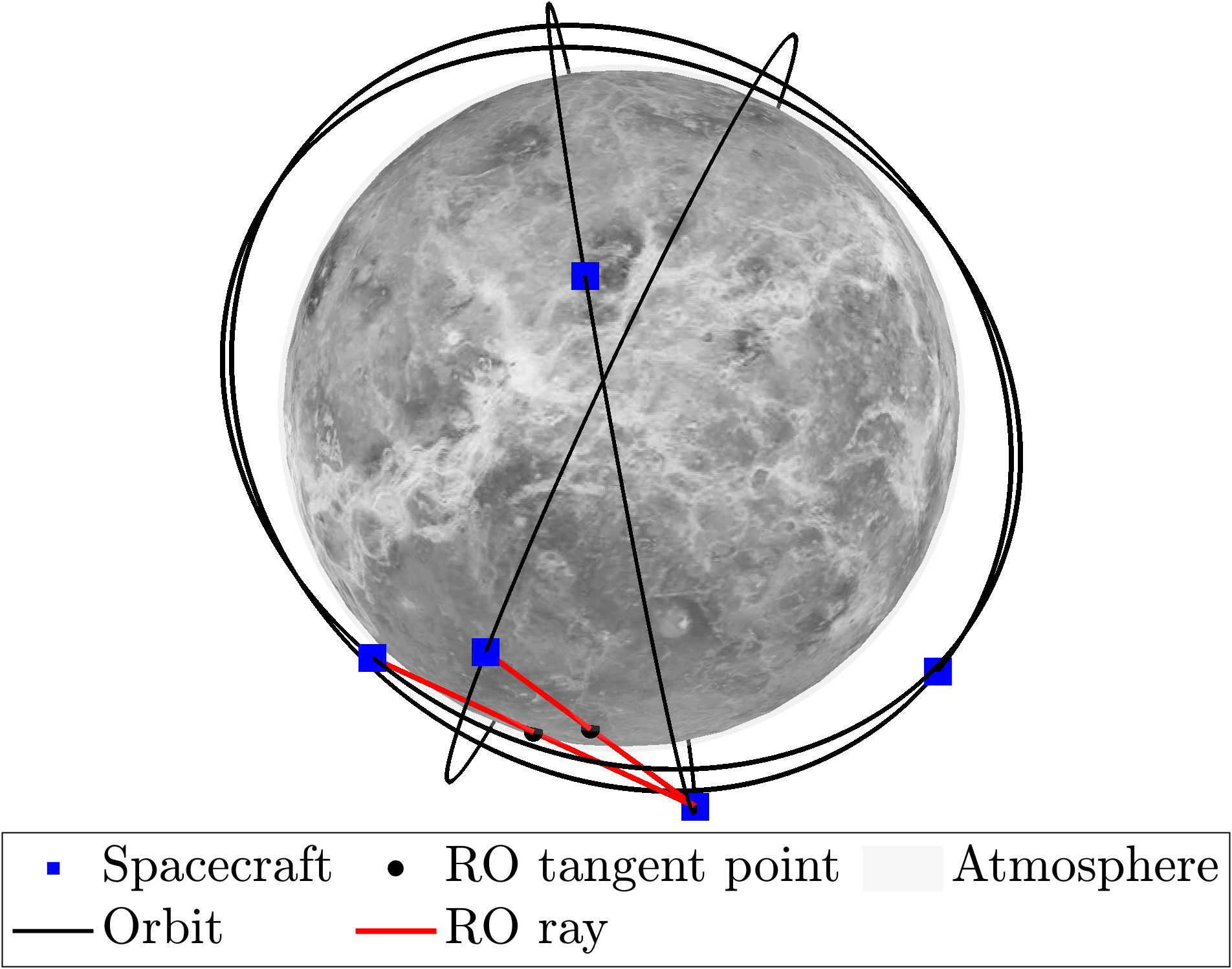}
        \caption{Constellation configuration.}
        \label{fig:location_3d}
    \end{subfigure}
    \begin{subfigure}[c]{0.492\linewidth}
        \centering
        \includegraphics[width=0.47\linewidth]{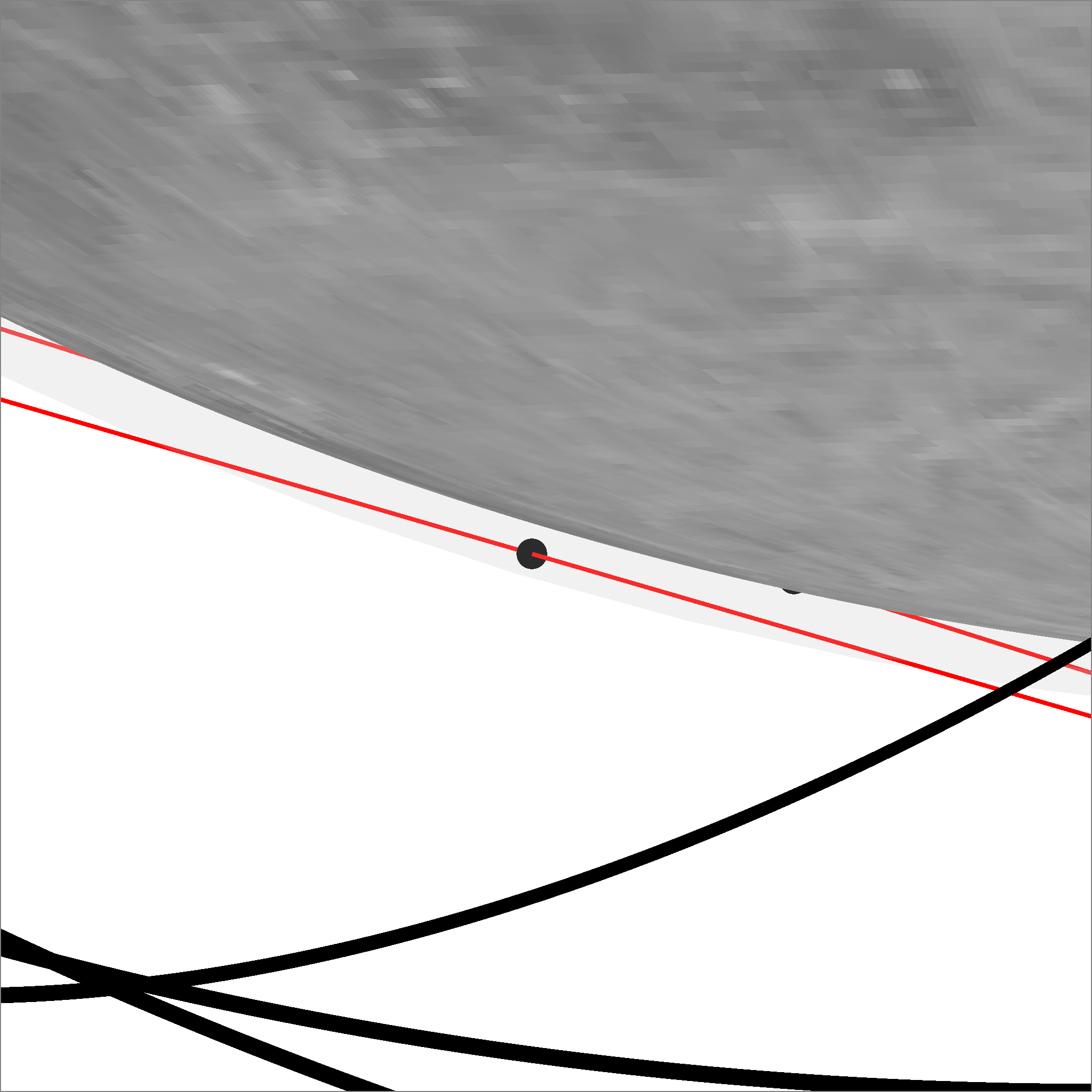}
        \caption{Detailed view of the RO tangent point.}
        \label{fig:location_3d_zoom}
    \end{subfigure}
    \caption{Optimal RO constellation.}
    \label{fig:case_study_location_3d}
\end{figure}

Table~\ref{table:numexp_cd_results_oe} presents the orbital elements corresponding to the optimal constellation configuration. All selected orbital slots share the same orbital radius of \SI{7358.22}{km}, but differ in their inclination, RAAN, and argument of latitude, yielding an asymmetric constellation configuration. Figure~\ref{fig:location_3d} illustrates the constellation configuration, the RO rays, and the location of the RO tangent points after \SI{720}{s} from the mission epoch. In addition, Fig.~\ref{fig:location_3d_zoom} presents a magnified view of the location of one RO tangent point. 
\begin{table}[htpb]
    \caption{Orbital elements for selected orbital slots.}
    \centering
    \label{table:numexp_cd_results_oe}%
    \begin{tabular}{ccccc}
        \hline
        \hline
        $j$ & Orbital radius, km & Inclination, deg & RAAN, deg & Argument of latitude, deg\\
        \hline
         281 & 7358.22 & 100 & 216 & 190 \\
        344 & 7358.22 & 115 & 144 & 250 \\
        347 & 7358.22 & 65 & 216 & 250 \\
        374 & 7358.22 & 115 & 288 & 250 \\
        431 & 7358.22 & 100 & 216 & 310 \\
        \hline
        \hline
    \end{tabular}
\end{table}

\begin{figure}[htpb]
    \centering
    \begin{subfigure}[c]{0.492\linewidth}
        \centering
        \includegraphics[width=0.95\linewidth]{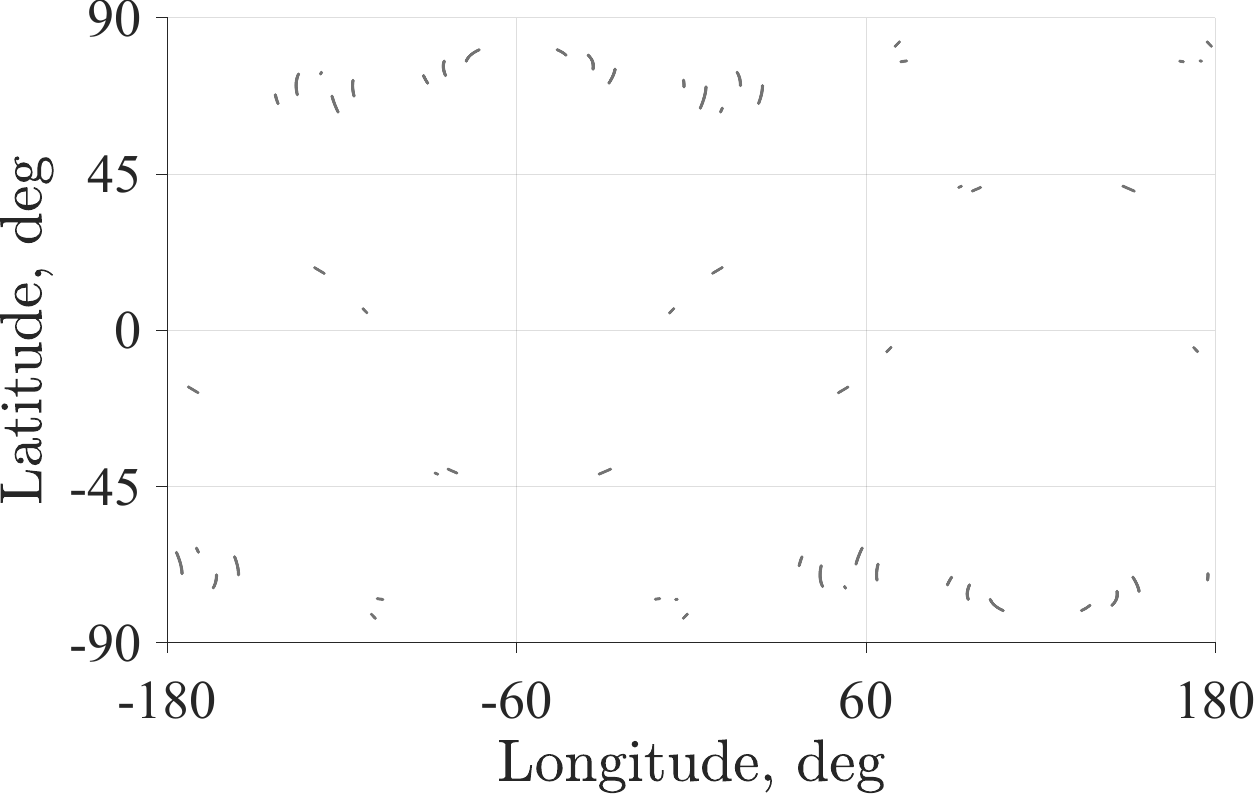}
        \caption{Optimization horizon.}
        \label{fig:location_tangent_short}
    \end{subfigure}
    \begin{subfigure}[c]{0.492\linewidth}
        \includegraphics[width=0.95\linewidth]{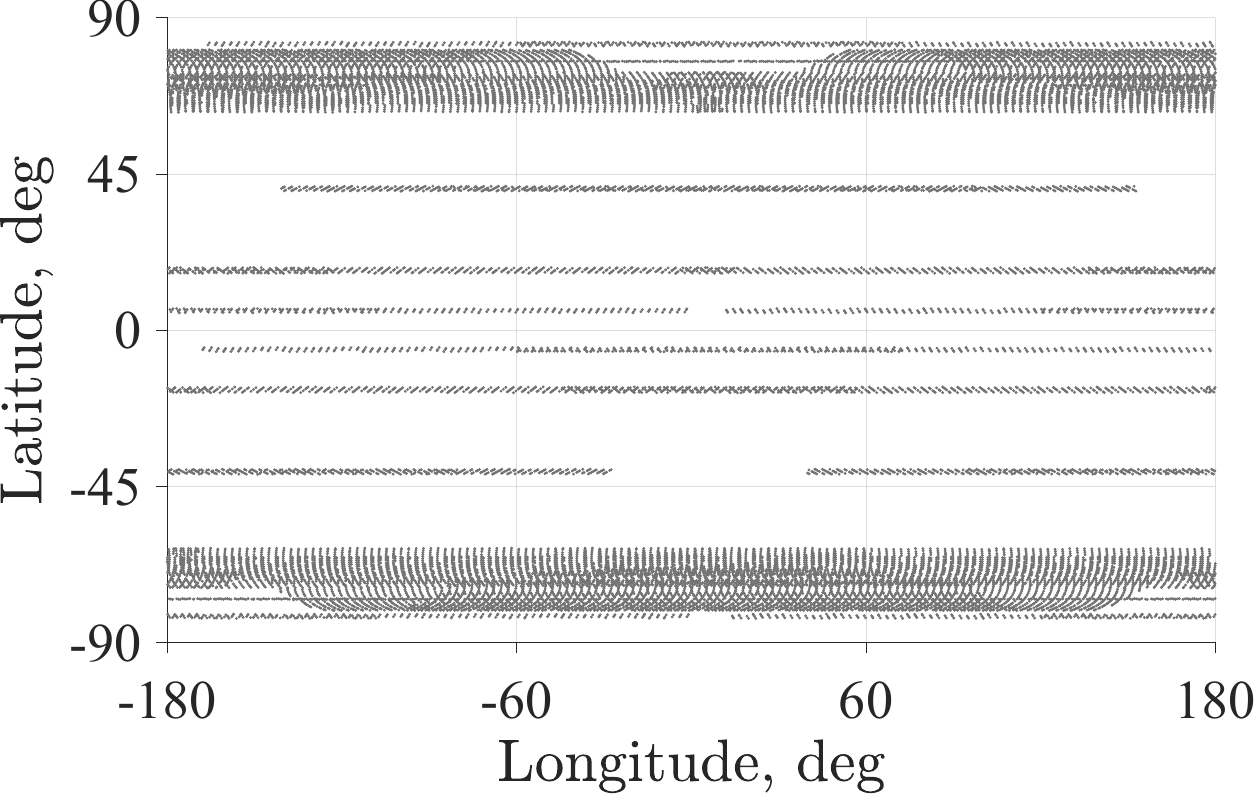}
    \caption{One Venusian day.}
    \label{fig:location_tangent_long}
    \end{subfigure}
    \caption{RO tangent point locations.}
    \label{fig:placeholder}
\end{figure}

Figure~\ref{fig:location_tangent_short} presents the location of the RO tangent points throughout the optimization horizon in the latitude-longitude plane. The bulk of these points is located in the higher-latitude regions, primarily caused by the near-polar inclination of the selected orbital slots. The lower-latitude regions still present RO tangent points but at a lower measurement frequency.

Figure~\ref{fig:location_tangent_long} shows the optimal constellation's RO tangent points obtained by simulating their motion for one Venusian day. During this period of time, the constellation obtains measurements across all longitude bands and the latitude bands defined approximately between $\pm80$ deg. Notably, these latitude bounds coincide with the maximum and minimum bounds the constellation achieves during the optimization horizon. Thus, simulating the constellation over longer time periods helps measure additional longitudes due to Venus' rotation but does not extend the measured latitudes. 

\subsubsection{Setup: RO-CRP}
Suppose volcanic activity is detected over the Ganiki Planitia quadrangle (25--50$\degree$N, 180--210$\degree$E \cite{ganiki_planitia}) after one Venusian day of the deployment of the RO-CDP constellation. Then, RO-CRP is used to optimally reconfigure the constellation and maximize RO measurement rewards over a \SI{24}{h} mission horizon, discretized in 480 time steps of size \SI{180}{s}. 

The Venusian atmosphere is uniformly discretized into 11 longitude cells, 10 latitude cells, and two altitude cells (with lower and upper bounds of 50 and \SI{90}{km}, respectively), yielding 220 atmospheric regions. During the first half of the mission horizon, we assign an RO measurement reward of one to the upper atmospheric regions and two to the lower atmospheric regions that encompass the Ganiki Planitia quadrangle. In a similar setting, during the second half of the mission horizon, RO measurement rewards of two and one are assigned to the upper and lower atmospheric regions of the Ganiki Planitia quadrangle, respectively. For the remaining atmospheric regions, we assign an RO measurement reward of zero throughout the entire mission horizon.

Each spacecraft has a randomly assigned $\Delta v$ transfer budget $\varepsilon_s$ of \SI{0.69}{km/s}, \SI{0.61}{km/s}, \SI{0.71}{km/s}, \SI{0.79}{km/s}, and \SI{0.68}{km/s}, and their orbital slots are generated as defined in Sec.~\ref{sec:numexp_benders_cr}. In particular, each spacecraft has three altitude steps (\textit{i.e.,} its current altitude, one \SI{500}{km} above, and one \SI{250}{km} below), 17 phasing steps, and a scaling factor of 0.6. Thus, each spacecraft has 225 transfer orbital slots, yielding an orbital slot set $\mathcal{J}$ of cardinality 1275, and a spacecraft pairs set $\mathcal{P}$ of cardinality \num{812175}.

\subsubsection{Results: RO-CRP}
The optimally reconfigured constellation obtains a total RO measurement reward of 90. Conversely, the baseline constellation (\textit{i.e.,} the constellation configuration from Sec.~\ref{sec:results_rocdp}) is unable to collect any reward. Table~\ref{table:numexp_cr_results_oe} presents the initial and final orbital slots and $\Delta v$ transfer budget $\varepsilon_s$ dedicated to maneuvering for each spacecraft. Neither of them fully uses their allocated $\Delta v$ transfer budget $\varepsilon_s$; among them, the minimum and maximum percentage usage of \SI{77.88}{\%} and \SI{96.02}{\%}  correspond to spacecrafts $s=1$ and $s=2$, respectively. This indicates that fully using the allocated $\Delta v$ transfer budget $\varepsilon_s$ does not yield larger objective values. Focusing on the orbital transfers, spacecraft $s=1$ executes a Hohmann transfer and phasing maneuvers; spacecrafts $s=2$ and $s=5$ perform a Hohmann transfer, RAAN change, and phasing maneuvers; spacecraft $s=3$ conducts RAAN and phasing maneuvers; and spacecraft $s=4$ performs inclination and phasing maneuvers.
\begin{table}[htpb]
    \caption{Final orbital elements after reconfiguration.}
    \centering
    \label{table:numexp_cr_results_oe}%
    \begin{tabular}{cccccccccc}
        \hline
        \hline
        \multirow{2}{*}{$s$} & \multicolumn{2}{c}{Orbital radius, km} & \multicolumn{2}{c}{Inclination, deg} &\multicolumn{2}{c}{RAAN, deg}& \multicolumn{2}{c}{Argument of latitude, deg}&\multirow{2}{*}{$\Delta v$, \si{km/s}}\\
        \cmidrule(lr){2-3} \cmidrule(lr){4-5} \cmidrule(lr){6-7} \cmidrule(lr){8-9}
        & Initial & Final & Initial & Final& Initial & Final& Initial & Final& \\
        \hline
         1 & 7358.22 & 7858.22 & 100 & 100    & 216  & 216    & 70.62  & 303.56 & 0.53\\
         2 & 7358.22 & 7858.22 & 115 & 115    & 144  & 146.16 & 130.62 & 151.79 & 0.58\\
         3 & 7358.22 & 7358.22 & 65  & 65     & 216  & 213.56 & 130.62 & 278.85 & 0.63\\
         4 & 7358.22 & 7358.22 & 115 & 119.08 & 288  & 288    & 130.62 & 24.74  & 0.72\\
         5 & 7358.22 & 7858.22 & 100 & 100    & 216  & 213.78 & 190.62 & 21.21  & 0.61\\
        \hline
        \hline
    \end{tabular}
\end{table}

\begin{figure}[htpb]
    \centering
    \includegraphics[width=0.41\linewidth]{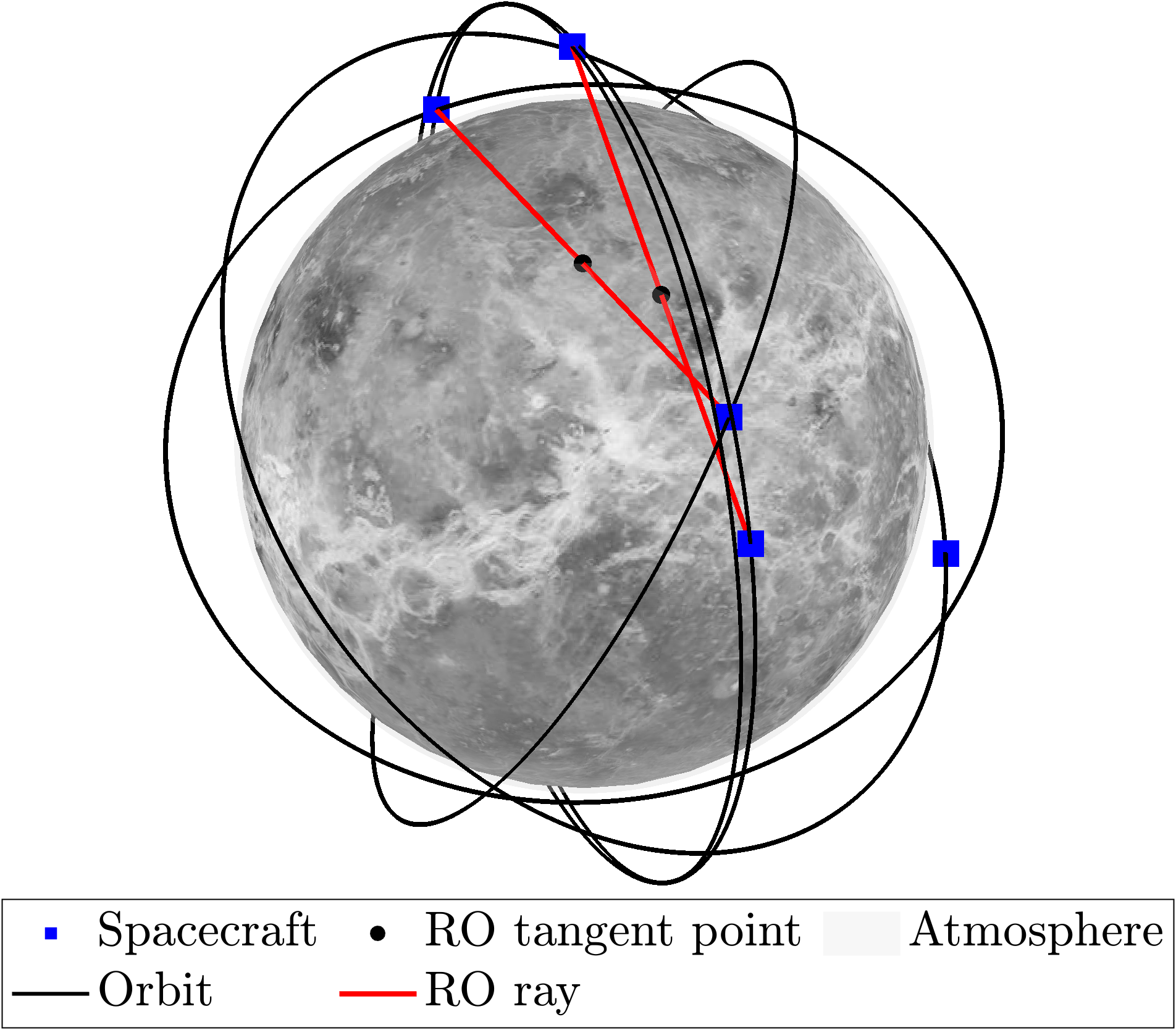}
    \caption{Optimal RO constellation configuration.}
    \label{fig:reconfiguration_3d}
\end{figure}

\begin{figure}[htpb]
    \centering
        \includegraphics[width=0.95\linewidth]{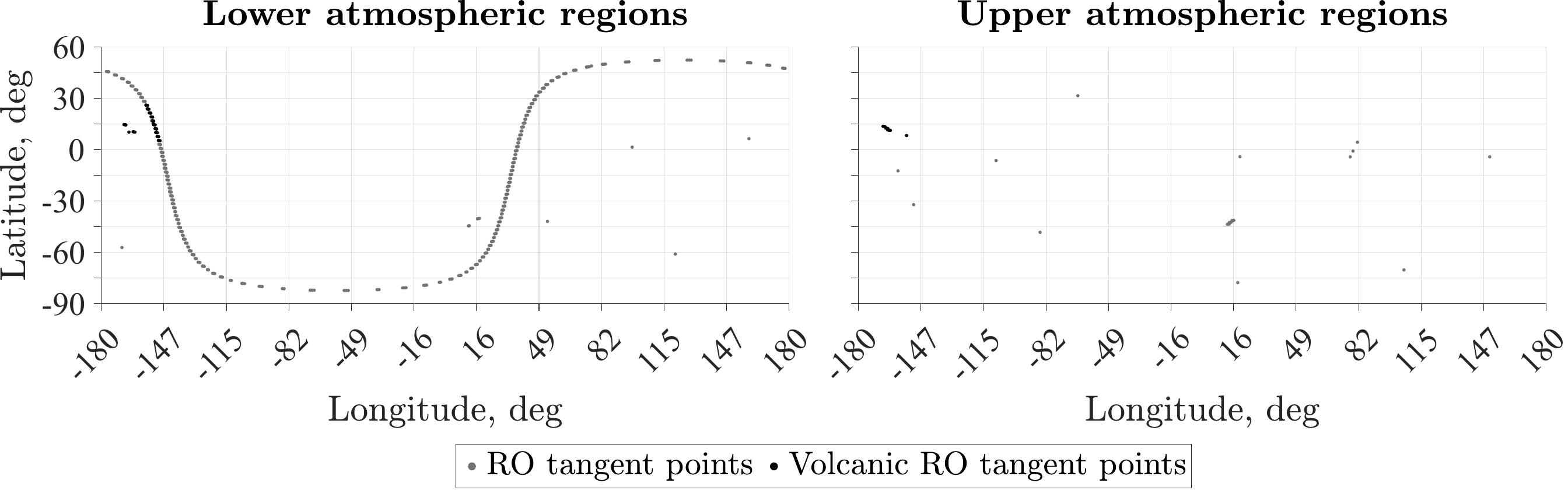}
        \caption{RO tangent point locations.}
        \label{fig:case_study_rec_tangent}
\end{figure}

Figure~\ref{fig:case_study_rec_tangent} shows the RO tangent point locations for the lower and upper atmospheric regions. The reconfigured constellation conducts 379 and 32 RO measurements over the lower and upper atmospheric regions. Among them, those colored black indicate that they measure an atmospheric region with volcanic activity and are associated with a time-varying RO measurement reward. Focusing on the distribution of RO tangent points, the reconfigured constellation performs 38 and 11 RO measurements over the lower and upper atmospheric regions, respectively, each associated with an RO measurement reward. 

This experiment sheds light on the benefits of constellation reconfiguration to address changes in the mission objectives. Overall, the initially designed constellation configuration cannot satisfy the new mission objectives; therefore, using RO-CRP, the constellation is reconfigured into a multi-layer, asymmetric constellation capable of measuring the atmospheric regions impacted by volcanic activity when desired.

\subsection{Sensitivity Analysis} \label{sec:mission_design}

This numerical experiment performs a sensitivity analysis of the optimal six-spacecraft constellation configuration with respect to a diverse set of complex, time-varying RO measurement rewards.

\subsubsection{Setup}
To test the sensitivity of the constellation configurations to complex RO measurement rewards, we propose five instances that each represent a unique mission scenario. All instances have an optimization horizon of \SI{96}{h} (\textit{i.e.,} four Earth days), uniformly discretized into time steps of size \SI{180}{s}. Furthermore, a common set of orbital slots is defined from all possible combinations of orbital elements defined in Table~\ref{table:numexp_md_params_oe}. Therefore, all instances share the same set of spacecraft pairs.

The first instance discretizes the Venusian atmosphere into three uniform latitude bands and four uniform longitude bands. This instance simulates a scanner that targets the first longitude band (\textit{i.e.,} the eastmost) on the first day and advances westward each day; that is, the second on the second day and the remaining bands accordingly. Therefore, we assign a reward equal to one to the atmospheric regions that lie inside the longitude band of interest at the time of interest and zero to all others. 

The remaining instances discretize the Venusian atmosphere into four latitude bands and three longitude bands, representing four distinct scanners that target the latitude bands. The second instance targets the two outer latitude bands during the first two days and the two inner bands during the last two days of the optimization horizon. Conversely, the third instance targets the inner latitude bands during the first two days, and the outer bands during the last two days. The fourth instance focuses on the first latitude band (\textit{i.e.,} the southernmost) during the first day and advances northwards for the remaining days; that is, the second band during the second day and the remaining bands accordingly. The fifth instance models a scanner similar to instance four; however, it starts from the northernmost latitude band and advances southwards. 

\begin{table}[ht]
\centering
    \caption{Orbital elements used to generate orbital slots.}
    \label{table:numexp_md_params_oe}%
    \begin{tabular}{lc}
        \hline
        \hline
        Orbital elements & Values\\
        \hline
        Orbital radius, km & $\{7051.80, 8071.20\}$\\
        Inclination, deg & $\{35, 55, 75, 95, 115\}$\\
        RAAN, deg & $\{0, 72, 144, 216, 288\}$ \\
        Argument of latitude, deg & $\{10, 82, 154, 226, 298\}$\\
        \hline
        \hline
    \end{tabular}
\end{table}

\subsubsection{Results}

The optimal constellation configurations, corresponding to each instance, obtain a total RO measurement reward of 299, 567, 566, 284, and 286, respectively. The second and third instances, which correspond to missions targeting the latitude bands for two Earth days, yield RO measurement rewards that approximately double those obtained by the remaining instances targeting a single altitude band for one Earth day. This trend indicates that more dynamic mission requirements (\textit{i.e.,} less time assigned to each band) increase mission complexity, thereby reducing the RO measurement rewards obtained. 

\begin{figure}[htbp]
    \centering
    \renewcommand{\arraystretch}{0.8}
    \begin{tabular}{cccc}
        \includegraphics[height=4cm]{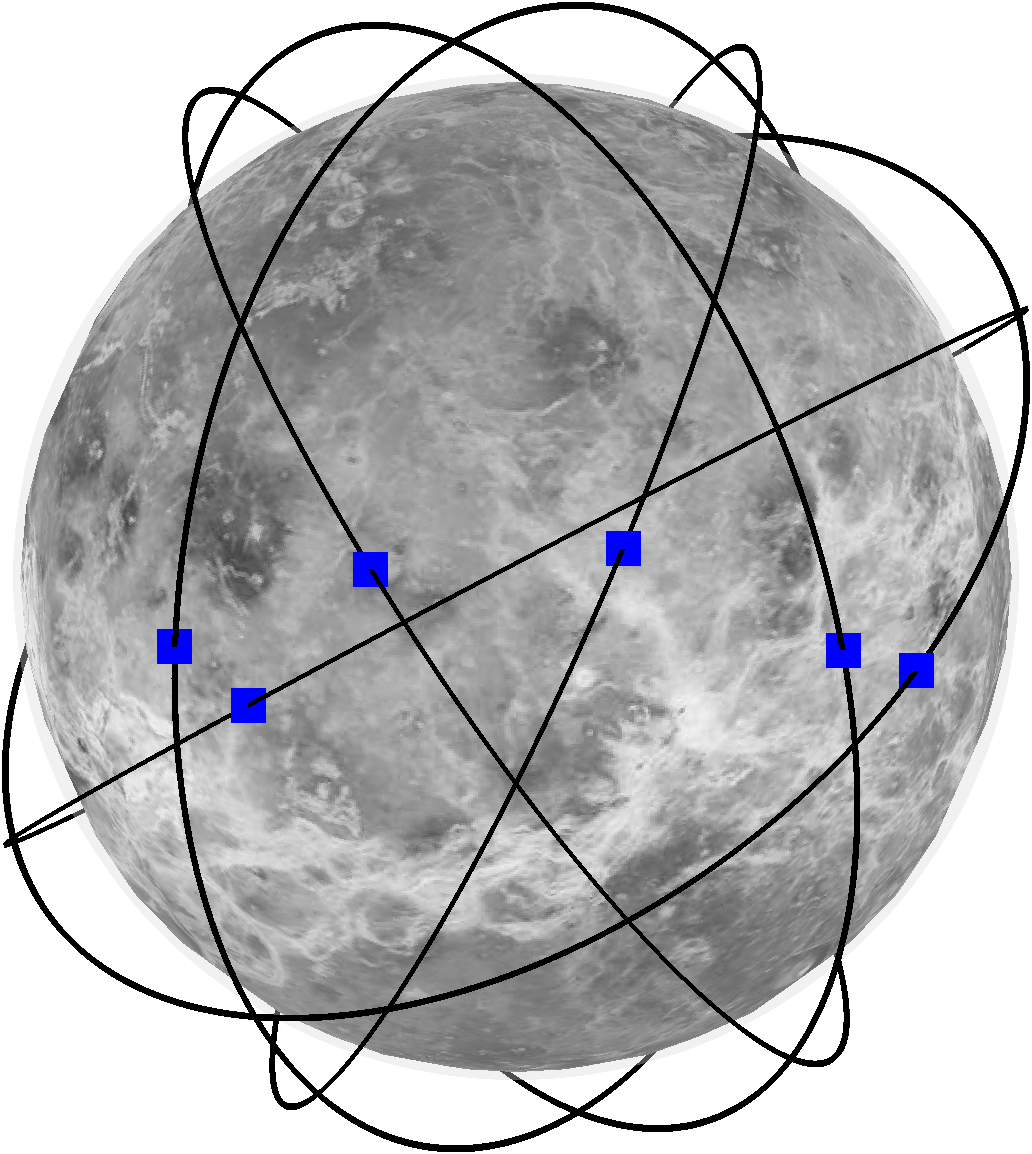} &

        \includegraphics[height=4cm]{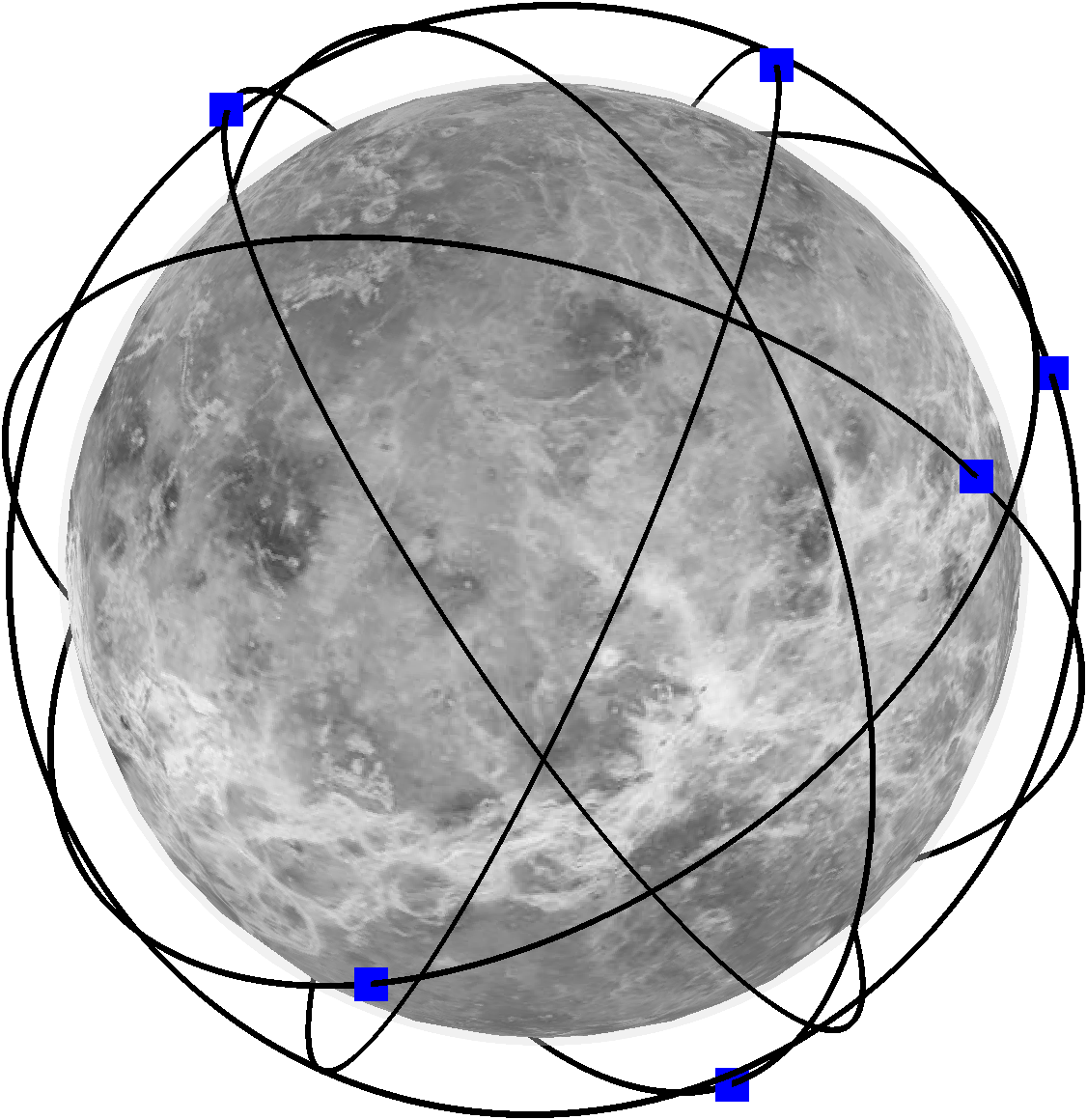} &

        \includegraphics[height=4cm]{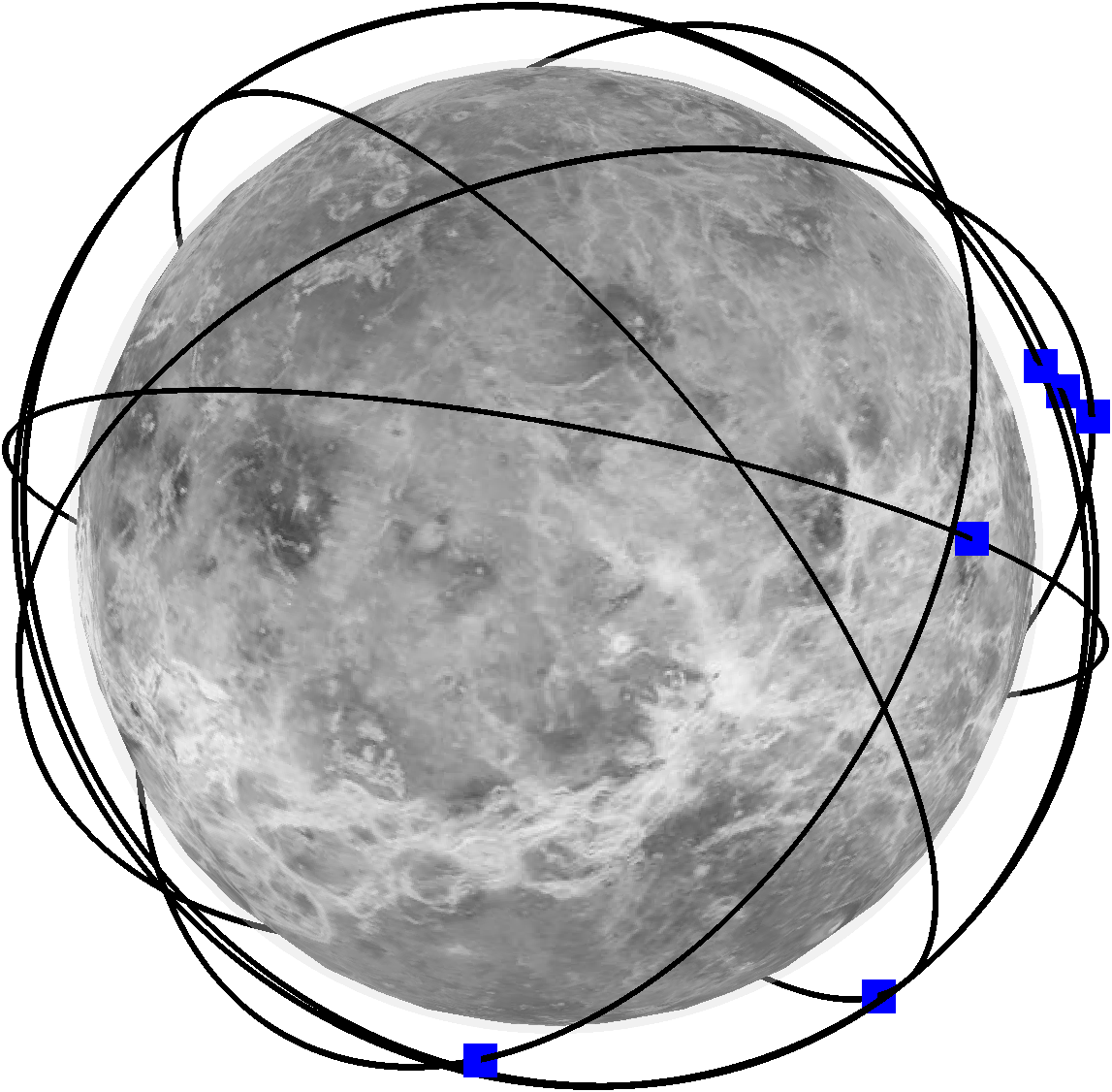} \\

        \textbf{(a)} First. & \textbf{(b)} Second. & \textbf{(c)} Third. \\
    \end{tabular}

    \begin{tabular}{cccc}
        \includegraphics[height=4cm]{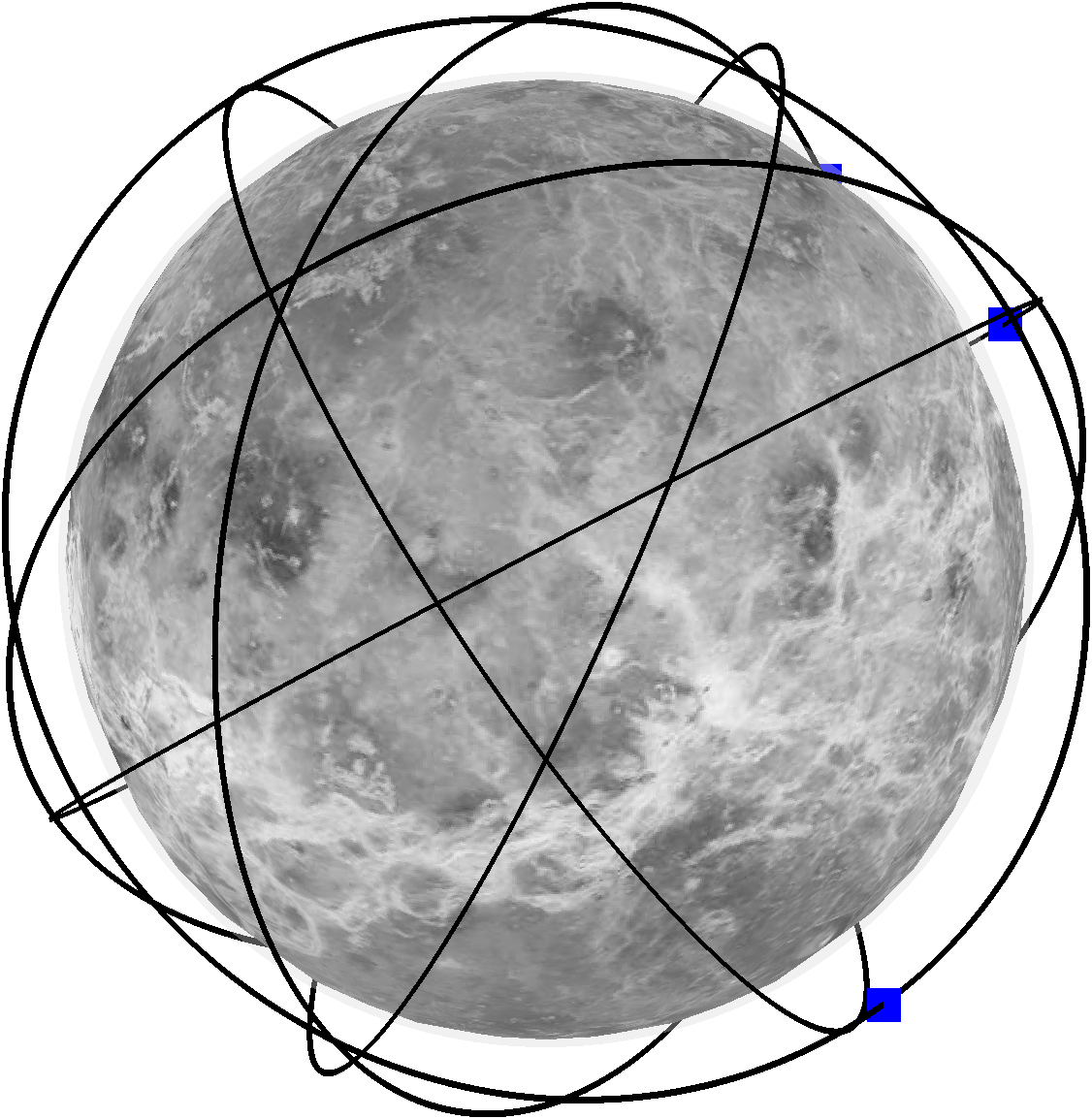} &

        \includegraphics[height=4cm]{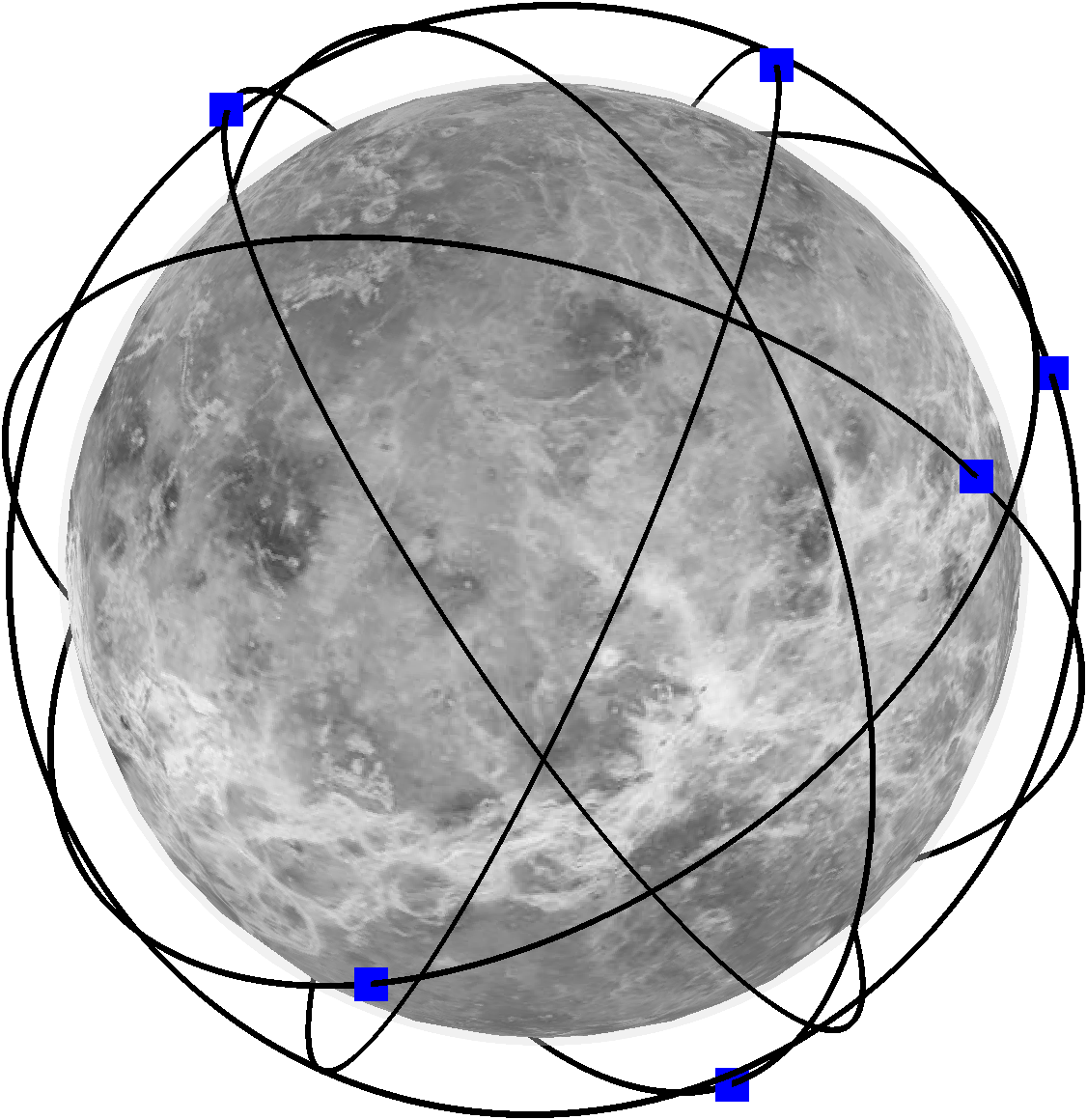} & \\
        \textbf{(d)} Fourth. & \textbf{(e)} Fifth. & \\
    \end{tabular}
    \caption{Constellation configurations corresponding to each instance.}
    \label{fig:mission_design_configurations}
\end{figure}

Figure~\ref{fig:mission_design_configurations} shows the optimal constellation configurations for each instance. All of them present configurations where no spacecraft share a common orbital plane, resulting in six unique orbital planes. Although all orbital planes have an orbital radius of \SI{7051.80}{km}, they differ in RAAN and inclination. Furthermore, it is particularly relevant that despite different RO measurement reward definitions, instances two and five share the same constellation configuration.

\begin{landscape}
    \begin{figure}[htbp]
    \centering
    \renewcommand{\arraystretch}{0.8}
    \begin{tabular}{cccc}
        \includegraphics[height=4.77cm]{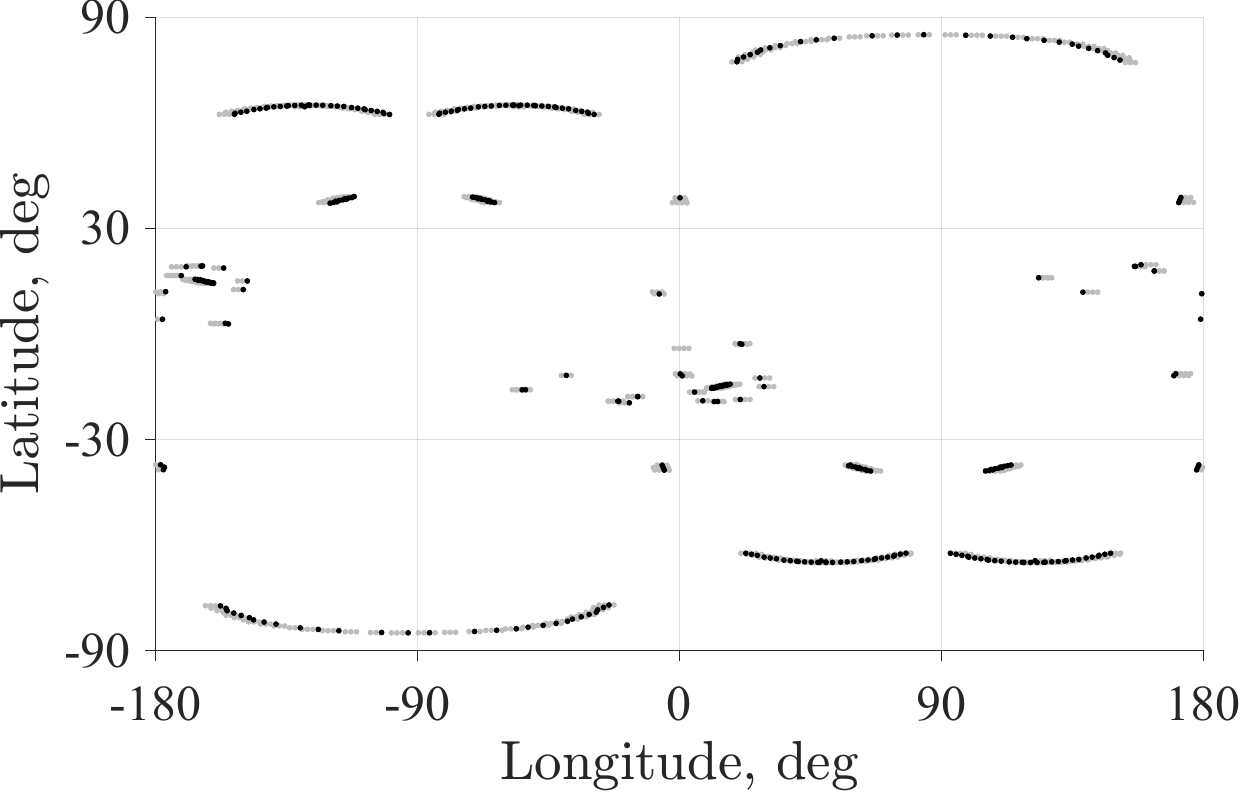} &

        \includegraphics[height=4.77cm]{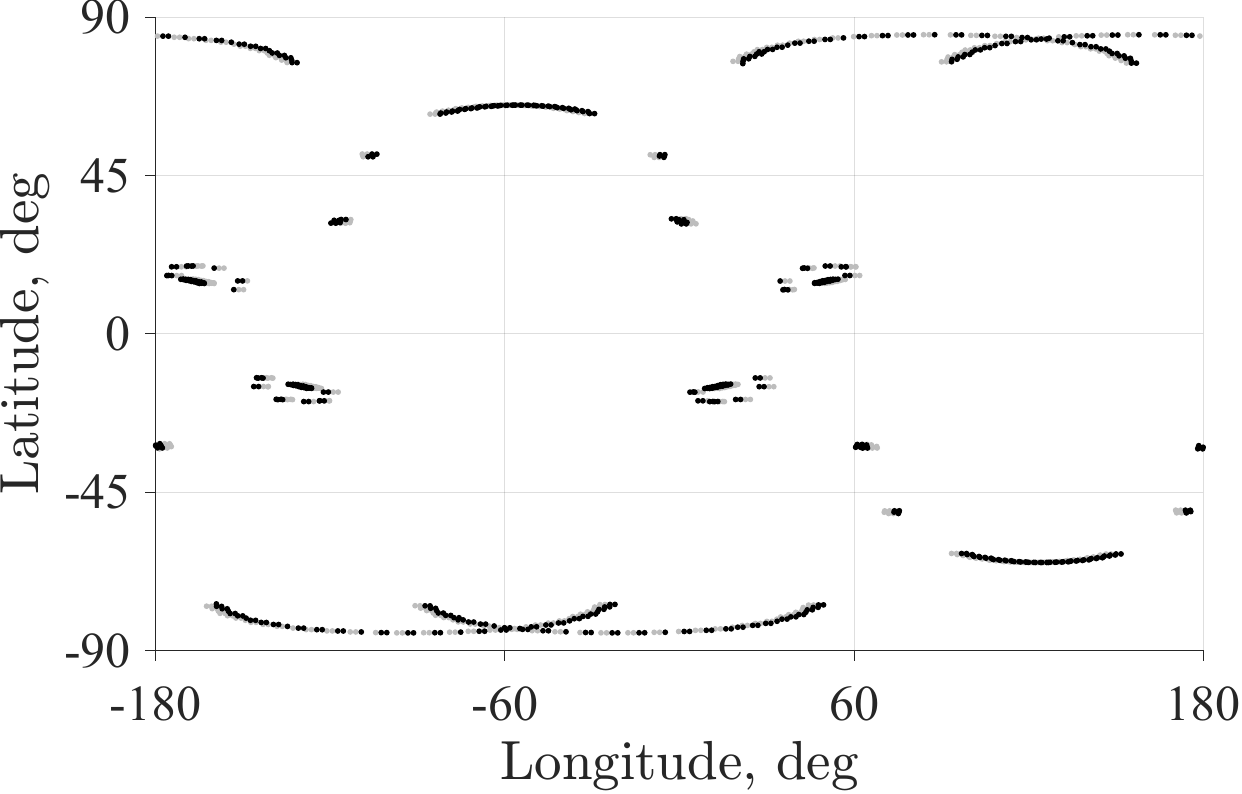} &

        \includegraphics[height=4.77cm]{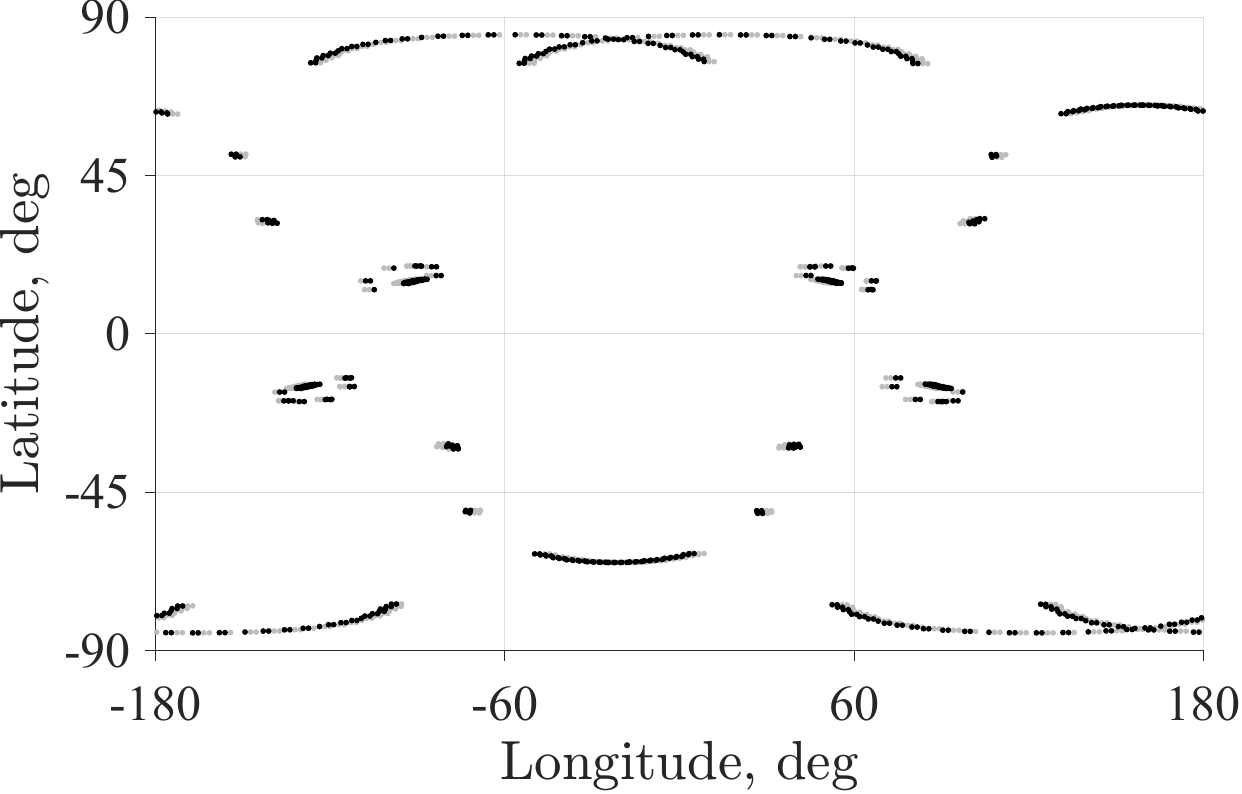} \\

        \textbf{(a)} First. & \textbf{(b)} Second. & \textbf{(c)} Third. \\
    \end{tabular}

    \begin{tabular}{cccc}
        \includegraphics[height=4.77cm]{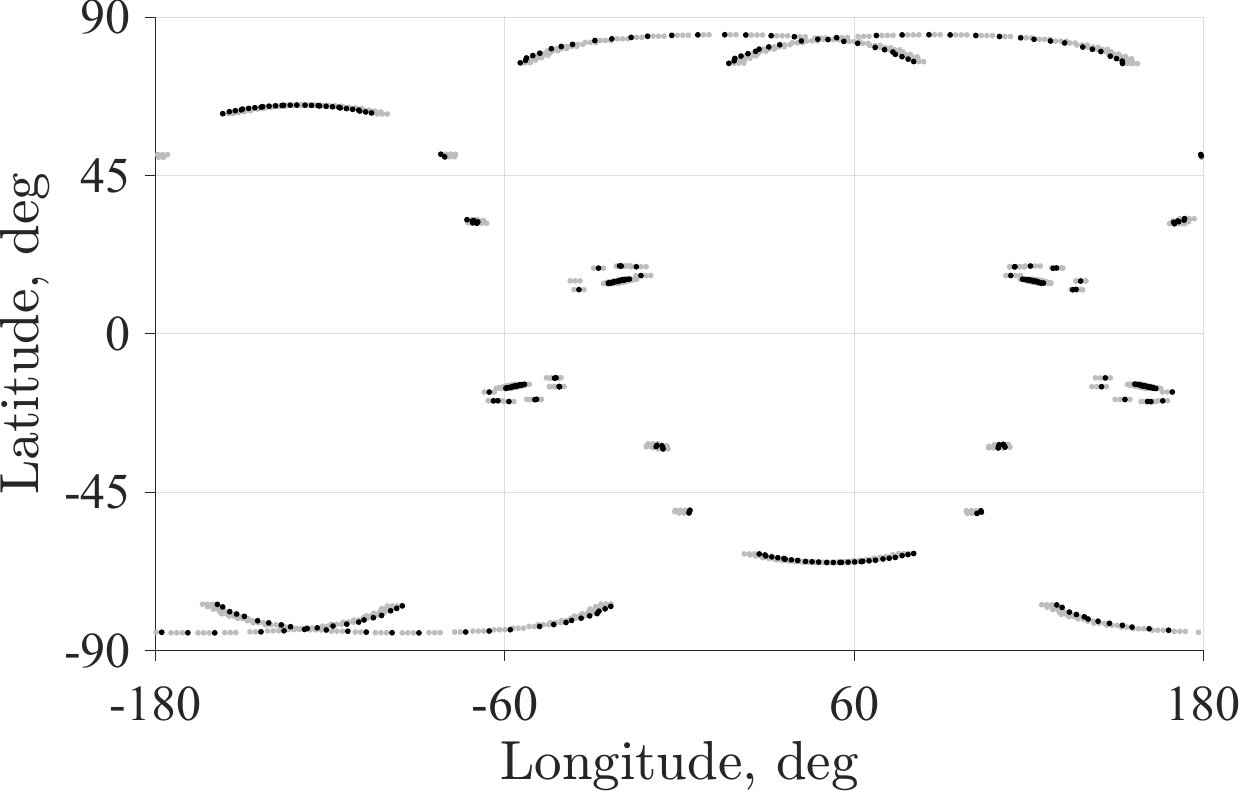} &

        \includegraphics[height=4.77cm]{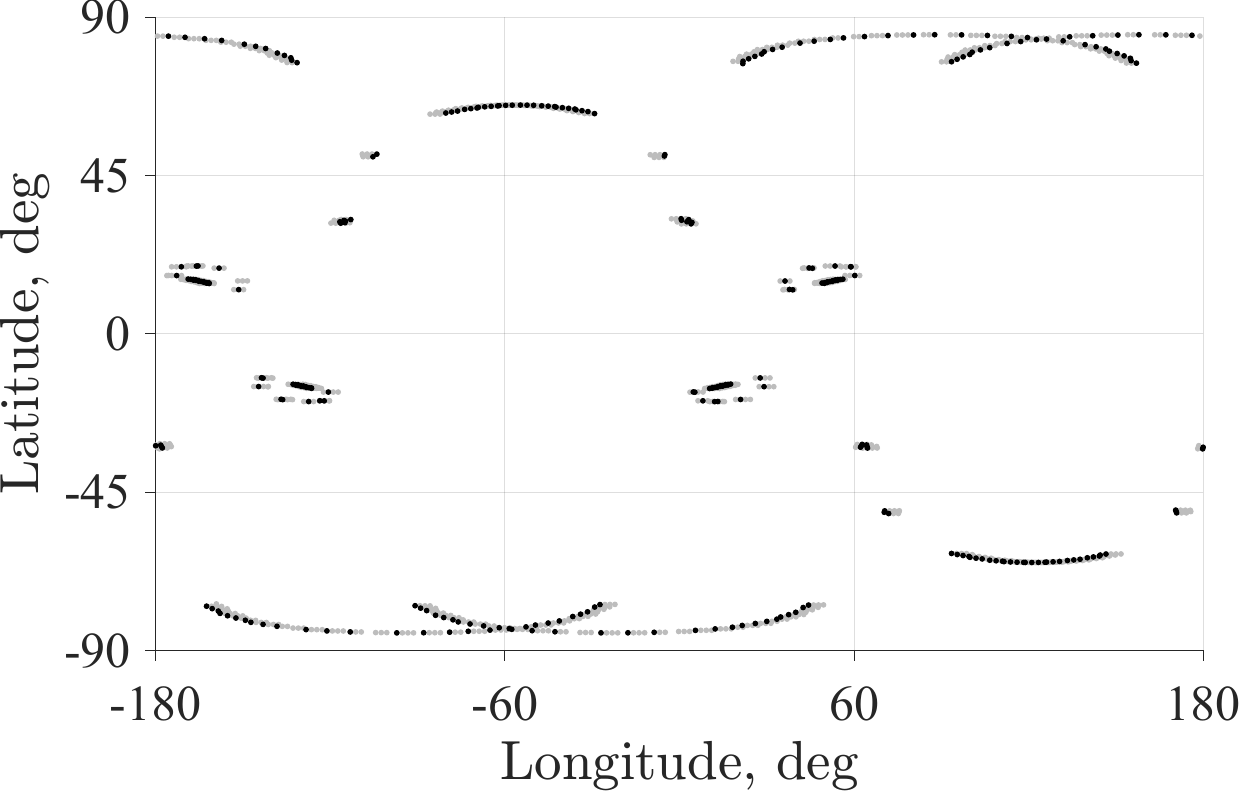} & \\
        \textbf{(d)} Fourth. & \textbf{(e)} Fifth. & \\
    \end{tabular}
    \caption{RO tangent point locations corresponding to each instance. Black RO tangent points have associated RO measurement rewards.}
    \label{fig:mission_design_tangent}
\end{figure}
\end{landscape}

Figure~\ref{fig:mission_design_tangent} displays the location of the RO tangent points for each instance, where black indicates those associated with a RO measurement reward equal to one. All instances coincide in that the bulk of the RO tangent points are black. This demonstrates RO-CDP's effectiveness in addressing complex mission narratives. Regarding instances two and five, which share the same constellation configuration and therefore have the same RO tangent point locations, there is an expected divergence in the distribution of black RO tangent points. Specifically, this divergence derives from the distinct definition of the RO measurement rewards.

\section{Conclusions}\label{sec:conclusions}
This article underpins the Venusian atmospheric RO mission design and operations pipeline by proposing RO-CDP and RO-CRP, formulated as ILPs and solved using a B\&BC algorithm. RO-CDP determines the optimal constellation configuration that maximizes RO measurement rewards, while RO-CRP maximizes these rewards through an optimal schedule of event-based constellation reconfigurations. Defining regional, time-dependent RO measurement rewards gives mission designers adaptability when handling complex mission narratives. 

A series of numerical experiments is presented throughout this article to demonstrate the problems' versatility and the benefits of solving them using B\&BC. Specifically, the first two numerical experiments show that adopting B\&BC to address RO-CDP and RO-CRP yields, in most instances, faster optimal solutions and, for suboptimal solutions, higher objective values, tighter upper bounds, or both. Furthermore, the problems' applicability and adaptability are demonstrated by addressing an RO mission design that initially tackles the entire Venusian atmosphere and then the mission objective changes towards sensing regional volcanic activity, requiring constellation reconfiguration. Lastly, we conduct sensitivity analyses by designing a set of constellation configurations to underpin different RO atmospheric scanning concepts (\textit{e.g.,} conducting RO atmospheric measurements over specific latitude or longitude bands during specific time periods), which directly show the flexibility provided by the proposed regional, time-dependent RO measurement rewards.

This article paves the way for future research efforts. For example, trajectory optimization methods may be investigated to determine low-thrust transfer maneuvers, potentially enabling less expensive constellation reconfigurations. In addition, the applicability of both problems for atmospheric studies on other planetary bodies may be explored. Lastly, RO-CRP may be reformulated as a multi-stage reconfiguration problem as outlined in Ref.~\cite{lee_jsr_2025}.

\appendix

\section*{Appendix A: Branch-and-Benders-Cut Benchmarking Instances}
Table~\ref{tab:instance_params_cd} outlines the key parameters for the 32 instances used in Sec.~\ref{sec:numexp_benders_cd}.
\begin{table}[htpb]
    \fontsize{9}{10}\selectfont
    \renewcommand{\arraystretch}{1.2}
    \caption{RO-CDP instances key parameters.}
    \label{tab:instance_params_cd}
    \centering
    \begin{subtable}{0.48\textwidth}
        \begin{tabular}{ccccccc|}
            \hline
            \hline
            Instance &$K$& $S$ & $J$ & $P$ & $O$ & Reward generation\\
            \hline
            1  & 100 & 5 & 216 & 23220 & 15 & uniform \\
            2  & 250 & 5 & 216 & 23220 & 15 & uniform \\
            3  & 100 & 6 & 216 & 23220 & 15 & uniform \\
            4  & 250 & 6 & 216 & 23220 & 15 & uniform \\
            5  & 100 & 5 & 216 & 23220 & 18 & uniform \\
            6  & 250 & 5 & 216 & 23220 & 18 & uniform \\
            7  & 100 & 6 & 216 & 23220 & 18 & uniform \\
            8  & 250 & 6 & 216 & 23220 & 18 & uniform \\
            9  & 100 & 5 & 360 & 64620 & 15 & uniform \\
            10 & 250 & 5 & 360 & 64620 & 15 & uniform \\
            11 & 100 & 6 & 360 & 64620 & 15 & uniform \\
            12 & 250 & 6 & 360 & 64620 & 15 & uniform \\
            13 & 100 & 5 & 360 & 64620 & 18 & uniform \\
            14 & 250 & 5 & 360 & 64620 & 18 & uniform \\
            15 & 100 & 6 & 360 & 64620 & 18 & uniform \\
            16 & 250 & 6 & 360 & 64620 & 18 & uniform \\
            \hline
            \hline
        \end{tabular}
    \end{subtable}
    \hfill
    \begin{subtable}{0.48\textwidth}
        \begin{tabular}{ccccccc}
            \hline
            \hline
            Instance &$K$& $S$ & $J$ & $P$ & $O$ & Reward generation\\
            \hline
            17 & 100 & 5 & 216 & 23220 & 15 & random \\
            18 & 250 & 5 & 216 & 23220 & 15 & random \\
            19 & 100 & 6 & 216 & 23220 & 15 & random \\
            20 & 250 & 6 & 216 & 23220 & 15 & random\\
            21 & 100 & 5 & 216 & 23220 & 18 & random \\
            22 & 250 & 5 & 216 & 23220 & 18 & random \\
            23 & 100 & 6 & 216 & 23220 & 18 & random \\
            24 & 250 & 6 & 216 & 23220 & 18 & random \\
            25 & 100 & 5 & 360 & 64620 & 15 & random \\
            26 & 250 & 5 & 360 & 64620 & 15 & random \\
            27 & 100 & 6 & 360 & 64620 & 15 & random \\
            28 & 250 & 6 & 360 & 64620 & 15 & random \\
            29 & 100 & 5 & 360 & 64620 & 18 & random \\
            30 & 250 & 5 & 360 & 64620 & 18 & random \\
            31 & 100 & 6 & 360 & 64620 & 18 & random \\
            32 & 250 & 6 & 360 & 64620 & 18 & random \\
            \hline
            \hline
        \end{tabular}
    \end{subtable}
\end{table}
Table~\ref{tab:instance_params_cr} presents the key parameters for the 18 instances considered in Sec.~\ref{sec:numexp_benders_cr}.
\begin{table}[htpb]
\fontsize{9}{10}\selectfont
    \renewcommand{\arraystretch}{1.2}
    \caption{RO-CRP instances key parameters.}
    \label{tab:instance_params_cr}
    \centering
        \begin{tabular}{ccccc|ccccccc}
            \hline
            \hline
              Instance &$K$& $J$ & $P$ & $O$ & Instance &$K$& $J$ & $P$ & $O$ \\
             \hline
                1  & 250 & 1125 & 632250  & 15 & 10 & 800 & 1500 & 1124250 & 30 \\
                2  & 800 & 1125 & 632250  & 15 & 11 & 250 & 1875 & 1756875 & 30 \\
                3  & 250 & 1500 & 1124250 & 15 & 12 & 800 & 1875 & 1756875 & 30 \\
                4  & 800 & 1500 & 1124250 & 15 & 13 & 250 & 1125 & 632250  & 45 \\
                5  & 250 & 1875 & 1756875 & 15 & 14 & 800 & 1125 & 632250  & 45 \\
                6  & 800 & 1875 & 1756875 & 15 & 15 & 250 & 1500 & 1124250 & 45 \\
                7  & 250 & 1125 & 632250  & 30 & 16 & 800 & 1500 & 1124250 & 45 \\
                8  & 800 & 1125 & 632250  & 30 & 17 & 250 & 1875 & 1756875 & 45 \\
                9  & 250 & 1500 & 1124250 & 30 & 18 & 800 & 1875 & 1756875 & 45 \\
             \hline
             \hline
        \end{tabular}
\end{table}
\clearpage
\section*{Appendix B: Plane Change Maneuvers}
The bounds of the plane change maneuvers are determined using the boundary value problems from Ref.~\cite{vallado}. Given spacecraft $s$ with a $\Delta v$ transfer budget $\varepsilon_{s}$, scaling factor $\beta_{s}$, and semi-major axis $a_{s}$, the change in inclination $\Delta inc_{s}$ is given as:
\begin{equation}
    \Delta inc_{s} = 2\beta_{s} \arcsin \left( \frac{\varepsilon_{s}}{2\sqrt{\mu_\text{Venus}/a_{s}}} \right)
\end{equation}
Then, given the initial inclination of spacecraft $s$ $inc_{s}$, the change in RAAN $\Delta \Omega_{s}$ is given as:
\begin{equation}
    \Delta \Omega_{s} = \beta_{s} \arccos \left( \frac{\cos{(\Delta inc_{s})}-\cos^2{(inc_{s})}}{\sin^2{(inc_{s})}} \right)
\end{equation}

\section*{Acknowledgment}
This work was supported in part by the NASA EPSCoR Rapid Response Research (R3) Program under award 80NSSC23M0168.

\bibliography{references}
\end{document}